\documentclass[11pt]{article}
\usepackage[letterpaper,hmargin=1in,vmargin=1.25in]{geometry}

\usepackage{url}

\usepackage{amsmath}
\usepackage{amssymb}

\usepackage{amsthm}

\theoremstyle{plain}
\newtheorem{theorem}{Theorem}[section]

\newtheorem{corollary}[theorem]{Corollary}

\newtheorem{definition}[theorem]{Definition}
\newtheorem{remark}[theorem]{Remark}

\newtheorem{replacements}[theorem]{Replacements}

\newcommand{\abs}[1]{\left\lvert#1\right\rvert}
\DeclareMathOperator{\Sim}{sim}
\DeclareMathOperator{\Dom}{dom}
\DeclareMathOperator*{\Limsup}{\varlimsup}

\newcommand{\N}{\mathbb{N}}
\newcommand{\Z}{\mathbb{Z}}
\newcommand{\Q}{\mathbb{Q}}
\newcommand{\R}{\mathbb{R}}
\newcommand{\X}{\{0,1\}^*}
\newcommand{\XI}{\{0,1\}^\infty}
\newcommand{\K}{H}

\newcommand{\noi}{\noindent}

\begin{document}


\begin{center}
{\Large \textbf{
  A statistical mechanical interpretation of\\
  algorithmic information theory
}}
\end{center}

\vspace{-2mm}

\begin{center}
Kohtaro Tadaki
\end{center}

\vspace{-5mm}

\begin{center}
Research and Development Initiative, Chuo University\\
1--13--27 Kasuga, Bunkyo-ku, Tokyo 112-8551, Japan\\
E-mail: tadaki@kc.chuo-u.ac.jp
\end{center}

\vspace{-2mm}

\begin{quotation}
\noi\textbf{Abstract.}
We develop a statistical mechanical interpretation of
algorithmic information theory
by introducing
the notion of
thermodynamic quantities, such as
free energy, energy, statistical mechanical entropy, and specific heat,
into algorithmic information theory.
We investigate the properties of these quantities
by means of program-size complexity
from the point of view of algorithmic randomness.
It is then discovered that, in the interpretation,
the temperature plays a role as the compression rate of
the values of all these thermodynamic quantities,
which include the temperature itself.
Reflecting this self-referential nature of
the compression rate of the temperature,
we obtain
fixed point theorems on compression rate.
\end{quotation}

\begin{quotation}
\noi\textit{Key words\/}:
algorithmic information theory,
algorithmic randomness,
Chaitin's $\Omega$,
compression rate,
fixed point theorem,
statistical mechanics,
temperature
\end{quotation}

\section{Introduction}

Algorithmic information theory is a framework to apply
information-theoretic and probabilistic ideas to recursive function theory.
One of the primary concepts of algorithmic information theory
is the \textit{program-size complexity}
(or \textit{Kolmogorov complexity}) $\K(s)$ of a finite binary string $s$,
which is defined as the length of the shortest binary
program
for the universal self-delimiting Turing machine $U$ to output $s$.
By the definition,
$\K(s)$ can be thought of as the information content of
the individual finite binary string $s$.
In fact,
algorithmic information theory has precisely the formal properties of
classical information theory (see \cite{C75}).
The concept of program-size complexity plays a crucial role in
characterizing the randomness of a finite or infinite binary string.
In \cite{C75} Chaitin introduced the halting probability $\Omega$
as an example of random infinite string.
His $\Omega$ is defined
as the probability that the universal self-delimiting Turing machine $U$ halts,
and plays a central role in the development of algorithmic information theory.
The first $n$ bits of the base-two expansion of $\Omega$ solves
the halting problem for a program of size not greater than $n$.
By this property,
the base-two expansion of $\Omega$ is shown to be
a random infinite binary string.
In \cite{C87a}
Chaitin encoded this random property of $\Omega$
onto an exponential Diophantine equation in the manner that
a certain property of the set of the solutions of the equation is
indistinguishable from coin tosses.
Moreover, based on this random property of the equation,
Chaitin derived several quantitative versions of
G\"odel's incompleteness theorems.

In \cite{T99,T02}
we generalized Chaitin's halting probability $\Omega$ to $\Omega^D$
by
\begin{equation}\label{intro-omegad}
  \Omega^D
  =
  \sum_{p\in\Dom U}2^{-\frac{\abs{p}}{D}},
\end{equation}
so that the degree of randomness of $\Omega^D$ can be controlled by
a real number $D$ with $0<D\le 1$.
Here, $\Dom U$ denotes the set of all programs $p$ for $U$.
As $D$ becomes larger, the degree of randomness of $\Omega^D$ increases.
When $D=1$,
$\Omega^D$ becomes a random real number, i.e., $\Omega^1=\Omega$.
The properties of $\Omega^D$ and its relations to self-similar sets
were studied in \cite{T99,T02}.

Recently, Calude and Stay \cite{CS07} pointed out
a formal correspondence between
$\Omega^D$ and a partition function in statistical mechanics.
In statistical mechanics,
the partition function $Z(T)$ at temperature $T$ is defined by
\begin{equation*}
  Z(T)=\sum_{x\in X}e^{-\frac{E_x}{kT}},
\end{equation*}
where $X$ is a complete set of energy eigenstates of a statistical mechanical system
and $E_x$ is the energy of an energy eigenstate $x$.
The constant $k$ is called the Boltzmann Constant.
The partition function $Z(T)$ is of particular importance
in equilibrium statistical mechanics.
This is because all the thermodynamic quantities of the system
can be expressed by using the partition function $Z(T)$,
and the knowledge of $Z(T)$ is sufficient to understand
all the macroscopic properties of the system.
Calude and Stay \cite{CS07} pointed out that
the partition function $Z(T)$ has the same form as $\Omega^D$
by performing the following replacements in $Z(T)$:
\begin{replacements}\label{CS07}\hfill
\begin{enumerate}
  \item Replace the complete set $X$ of energy eigenstates $x$
    by the set $\Dom U$ of all programs $p$ for $U$.
  \item Replace the energy $E_x$ of an energy eigenstate $x$
    by the length $\abs{p}$ of a program $p$.
  \item Set the Boltzmann Constant $k$ to $1/\ln 2$,
    where the $\ln$ denotes the natural logarithm.
\end{enumerate}
\end{replacements}
In this paper,
inspired by their suggestion,
we develop a statistical mechanical interpretation of
algorithmic information theory,
where $\Omega^D$ appears as a partition function.

Generally speaking,
in order to give a statistical mechanical interpretation to a framework
which looks unrelated to statistical mechanics at first glance,
it is important to identify a microcanonical ensemble
in the framework.
Once we can do so,
we can easily develop an equilibrium statistical mechanics on the framework
according to the theoretical development of
normal equilibrium statistical mechanics.
Here, the microcanonical ensemble is a certain sort of
uniform probability distribution.
In fact,
in the work \cite{T07}
we developed
a statistical mechanical interpretation of
the noiseless source coding scheme
in information theory
by identifying a microcanonical ensemble in the scheme.
Then, in \cite{T07} the notions in statistical mechanics
such as statistical mechanical entropy, temperature,
and thermal equilibrium are translated into the context of
noiseless source coding.

Thus,
in order to develop a statistical mechanical interpretation of
algorithmic information theory,
it is appropriate to identify a microcanonical ensemble in the framework of
the theory.
Note, however, that
algorithmic information theory is not a physical theory
but a purely mathematical theory.
Therefore,
in order to obtain significant results
for the development of algorithmic information theory itself,
we have to develop
a statistical mechanical interpretation
of algorithmic information theory
in a mathematically rigorous manner,
unlike in normal statistical mechanics in physics
where arguments are not necessarily mathematically rigorous.
A fully rigorous mathematical treatment of statistical mechanics is
already developed (see \cite{Ru99}).
At present,
however,
it would not as yet seem to be an easy task to merge
algorithmic information theory
with this mathematical treatment in a satisfactory manner.

On the other hand,
if we do not stick to the mathematical strictness of an argument,
we can develop a statistical mechanical interpretation of
algorithmic information theory
while realizing a perfect correspondence to normal statistical mechanics.
In fact,
in the last part of this paper (i.e., in Section \ref{smi})
we develop a statistical mechanical interpretation of
algorithmic information theory
by making
an
argument
on the same level of mathematical strictness as statistical mechanics
in physics.
There,
we identify a microcanonical ensemble in algorithmic information theory
in a similar manner to \cite{T07},
based on the probability measure which gives Chaitin's $\Omega$
the meaning of the halting probability
actually.
In consequence,
for example,
the statistical mechanical meaning of $\Omega^D$ is clarified.

In the main part of this paper,
for mathematical strictness
we develop a statistical mechanical interpretation of
algorithmic information theory
in a different way from
the physical argument
in Section $\ref{smi}$.%
\footnote{
We make an argument in a fully mathematically rigorous manner
in this paper except for
Section \ref{smi}.
Any consequence of the argument in Section \ref{smi}
is not used
in any other parts of this paper.
}
We introduce
the notion of thermodynamic quantities
into algorithmic information theory
based on Replacements~\ref{CS07} above.
Section \ref{smi} plays a role
in clarifying
the statistical mechanical meaning of these notion
and motivating the introduction of them in the main part of this paper.

After the preliminary section on the mathematical notion needed in this paper,
we prove some results on the degree of randomness of real numbers
in Section \ref{Chaitin D-randomness}.
These results themselves and
the techniques used in proving these results
are frequently used throughout this paper.

Then,
in Section \ref{tcr}
we introduce
the notion of the thermodynamic quantities at any given fixed temperature $T$,
such as partition function, free energy, energy,
statistical mechanical entropy, and specific heat,
into algorithmic information theory
by performing Replacements~\ref{CS07}
for the corresponding thermodynamic quantities
in statistical mechanics.
These thermodynamic quantities in algorithmic information theory
are real numbers which depend only on the temperature $T$.
We prove that if the temperature $T$ is a computable real number with $0<T<1$
then, for each of these thermodynamic quantities,
the compression rate by the program-size complexity $H$ is equal to $T$.
Thus, the temperature $T$ plays a role as
the compression rate of the thermodynamic quantities
in this statistical mechanical interpretation
of algorithmic information theory.

Among all thermodynamic quantities
in thermodynamics,
one of the most typical thermodynamic quantities is temperature
itself.
Thus, based on the results of Section \ref{tcr},
the following question naturally arises:
Can the compression rate of the temperature $T$
be equal to the temperature itself
in the statistical mechanical interpretation
of algorithmic information theory~?
This question is rather self-referential.
However,
in Section \ref{fpt}
we answer it affirmatively
by proving Theorem~\ref{main}.
One consequence of Theorem~\ref{main} has the following form:
For every $T\in(0,1)$,
if $\Omega^T=\sum_{p\in\Dom U}2^{-\frac{\abs{p}}{T}}$
is a computable real number, then
\begin{equation*}
  \lim_{n\to\infty}\frac{H(T_n)}{n}=T,
\end{equation*}
where $T_n$ is the first $n$ bits of the base-two expansion of $T$.
This is just a fixed point theorem on compression rate,
which reflects the self-referential nature of the question.

The works \cite{T99,T02} on $\Omega^D$ might be regarded as
an elaboration of the technique used by Chaitin \cite{C75}
to prove that $\Omega$ is random.
The
mathematical
results
of this paper,
which are obtained except for in Section \ref{smi},
may be regarded as further elaborations of the technique.
%

Finally,
in Section \ref{smi},
based on a physical and informal argument
we develop a
total
statistical mechanical interpretation
of algorithmic information theory
which
attains
a perfect correspondence to normal statistical mechanics.
In consequence,
we justify the interpretation of $\Omega^D$
as a partition function
and clarify the statistical mechanical meaning of
the thermodynamic quantities introduced
into algorithmic information theory
in Section \ref{tcr}.

\section{Preliminaries}
\label{preliminaries}


We start with some notation about numbers and strings
which will be used in this paper.
$\N=\left\{0,1,2,3,\dotsc\right\}$ is the set of natural numbers,
and $\N^+$ is the set of positive integers.
$\Z$ is the set of integers, and
$\Q$ is the set of rational numbers.
$\R$ is the set of real numbers.
$\X=
\left\{
  \lambda,0,1,00,01,10,11,000,001,010,\dotsc
\right\}$
is the set of finite binary strings
where $\lambda$ denotes the \textit{empty string}.
For any $s \in \X$, $\abs{s}$ is the \textit{length} of $s$.
A subset $S$ of $\X$ is called a \textit{prefix-free set}
if no string in $S$ is a prefix of another string in $S$.
$\XI$ is the set of infinite binary strings,
where an infinite binary string is
infinite to the right but finite to the left.
For any $\alpha\in \XI$, $\alpha_n$ is the prefix of $\alpha$ of length $n$.
For any partial function $f$,
the domain of definition of $f$ is denoted by $\Dom f$.
We write ``r.e.'' instead of ``recursively enumerable.''

Normally, $o(n)$ denotes any one function $f\colon \N^+\to\R$ such
that $\lim_{n \to \infty}f(n)/n=0$.
On the other hand,
$O(1)$ denotes any one function $g\colon \N^+\to\R$ such that
there is $C\in\R$ with the property that
$\abs{g(n)}\le C$ for all $n\in\N^+$.

Let $T$ be an arbitrary real number.
$T\bmod 1$ denotes $T - \lfloor T \rfloor$,
where $\lfloor T \rfloor$ is the greatest integer less than or equal to $T$,
and $T \bmod' 1$ denotes $T - \lceil T \rceil + 1$,
where $\lceil T \rceil$ is the smallest integer greater than or equal to $T$.
Hence, $T\bmod 1 \in [0,1)$ but $T\bmod' 1 \in (0,1]$.
We identify a real number $T$ with
the infinite binary string $\alpha$ such that
$0.\alpha$ is the base-two expansion of $T\bmod 1$ with infinitely many zeros.
Thus, $T_n$ denotes the first $n$ bits of
the base-two expansion of $T\bmod 1$ with infinitely many zeros.

We say that a real number $T$ is \textit{computable} if
there exists a total recursive function $f\colon\N^+ \to \Q$ such that
$\abs{T-f(n)} < 1/n$ for all $n\in\N^+$.
We say that $T$ is \textit{right-computable} if
there exists a total recursive function $g\colon\N^+\to\Q$ such that
$T\le g(n)$ for all $n\in\N^+$ and $\lim_{n\to\infty} g(n)=T$.
We say that $T$ is \textit{left-computable} if $-T$ is right-computable.
%
It is then easy to see that,
for any $T\in\R$,
$T$ is computable if and only if
$T$ is both right-computable and left-computable.
See e.g.~\cite{PR89,W00}
for the detail of the treatment of
the computability of real numbers and real functions on a discrete set.

\subsection{Algorithmic information theory}
\label{ait}

In the following
we concisely review some definitions and results of
algorithmic information theory
\cite{C75,C87b}.
A \textit{computer} is a partial recursive function
$C\colon \X\to \X$
such that
$\Dom C$ is a prefix-free set.
For each computer $C$ and each $s \in \X$,
$\K_C(s)$ is defined by
$\K_C(s) =
\min
\left\{\,
  \abs{p}\,\big|\;p \in \X\>\&\>C(p)=s
\,\right\}$.
A computer $U$ is said to be \textit{optimal} if
for each computer $C$ there exists a constant $\Sim(C)$
with the following property;
if $C(p)$ is defined, then there is a $p'$ for which
$U(p')=C(p)$ and $\abs{p'}\le\abs{p}+\Sim(C)$.
It is then shown that there exists
an optimal computer.
We choose any one optimal computer $U$ as the standard one for use,
and define $\K(s)$ as $\K_U(s)$,
which is referred to as
the \textit{program-size complexity} of $s$,
the \textit{information content} of $s$, or
the \textit{Kolmogorov complexity} of $s$
\cite{G74,L74,C75}.
Thus, $H(s)$ has the following property:
\begin{equation}
  \forall\,C:\text{computer}\quad
  H(s) \le H_C(s) + \Sim(C). \label{eq: k}
\end{equation}
It can be shown that there is $c\in\N$ such that,
for any $s \neq \lambda$,
\begin{equation}
  H(s)\le\abs{s}+2\log_2\abs{s}+c. \label{eq: fas}
\end{equation}

For each $s\in \X$, $P(s)$ is defined as $\sum_{U(p)=s}2^{-\abs{p}}$.
\textit{Chaitin's halting probability} $\Omega$ is defined by
\begin{equation*}
  \Omega=\sum_{p\in\Dom U}2^{-\abs{p}}.
\end{equation*}
For any $\alpha\in\XI$,
we say that $\alpha$ is
\textit{weakly Chaitin random}
if there exists $c\in\N$ such that
$n-c\le \K(\alpha_n)$ for all $n\in\N^+$
\cite{C75,C87b}.
Then \cite{C75} showed that $\Omega$ is weakly Chaitin random.
For any $\alpha\in\XI$,
we say that $\alpha$ is
\textit{Chaitin random}
if $\lim_{n\to \infty} \K(\alpha_n)-n=\infty$ \cite{C75,C87b}.
It is then shown that,
for any $\alpha\in\XI$,
$\alpha$ is weakly Chaitin random if and only if
$\alpha$ is Chaitin random
(see \cite{C87b} for the proof and historical detail).
Thus $\Omega$ is Chaitin random.

The class of computers is equal to the class of functions
which are computed by \textit{self-delimiting Turing machines}.
A self-delimiting Turing machine is a deterministic Turing machine
which has two tapes, a program tape and a work tape.
The program tape is infinite to the right,
while the work tape is infinite in both directions.
An input string in $\X$ is put on the program tape.
See Chaitin \cite{C75} for the detail of self-delimiting Turing machines.
Let $M$ be a
self-delimiting Turing machine which
computes the optimal computer $U$.
Then $P(s)$ is the probability that $M$ halts and outputs $s$
when $M$ starts on the program tape filled with an infinite binary string
generated by infinitely repeated tosses of a fair coin.
Therefore $\Omega=\sum_{s\in\X} P(s)$ is the probability that
$M$ just halts
under the same setting.

The program-size complexity $\K(s)$ is originally defined
using the concept of program-size, as stated above.
However,
it is possible to define $\K(s)$ without referring to such a concept,
i.e.,
we first introduce a \textit{universal probability} $m$,
and then define $\K(s)$ as $-\log_2 m(s)$.
A universal probability is defined through the following two definitions
\cite{ZL70}.

\begin{definition}\label{lcsp}
  For any $r\colon \X\to[0,1]$,
  we say that $r$ is a \textit{lower-computable semi-measure} if
  $r$ satisfies the following two conditions:
  \begin{enumerate}
    \item $\sum_{s\in \X}r(s)\le 1$.
    \item There exists a total recursive function
    $f\colon\N^+\times \X\to\Q$
      such that, for each $s\in \X$,
      $\lim_{n\to\infty} f(n,s)=r(s)$ and
      $\forall\,n\in\N^+\;\,f(n,s)\le r(s)$.\qed
  \end{enumerate}
\end{definition}

\begin{definition}
  Let $m$ be a lower-computable semi-measure.
  We say that $m$ is a \textit{universal probability} if
  for any lower-computable semi-measure $r$,
  there exists a real number $c>0$ such that,
  for all $s\in \X$, $c\,r(s)\le m(s)$.\qed
\end{definition}

Chaitin \cite{C75} showed the following theorem.

\begin{theorem}\label{eup}
  Both $2^{-\K(s)}$ and $P(s)$ are universal probabilities.
  \qed
\end{theorem}

By Theorem \ref{eup}, we see that, for any universal probability $m$,
\begin{equation}\label{eq: K_m}
  \K(s)=-\log_2 m(s)+O(1).
\end{equation}
Thus it is possible to define $\K(s)$ as $-\log_2 m(s)$
with any one universal probability $m$ instead of as $\K_U(s)$.
Note that
the difference up to an additive constant is inessential to
algorithmic information theory.

In the works \cite{T99,T02},
we generalized the notion of
the randomness of an infinite binary string
so that the degree of the randomness can be characterized
by a real number $D$ with $0<D\le 1$ as follows.

\begin{definition}[weakly Chaitin $D$-random]
  Let $D\in\R$ with $D\ge 0$,
  and let $\alpha \in\XI$.
  We say that $\alpha$ is \textit{weakly Chaitin $D$-random} if
  there exists $c\in\R$ such that
  $Dn-c \le H(\alpha_n)$
  for all $n\in\N^+$.
  \qed
\end{definition}

\begin{definition}[$D$-compressible]
Let $D\in\R$ with $D\ge 0$,
and let $\alpha \in\XI$.
We say that $\alpha$ is \textit{$D$-compressible} if
$H(\alpha_n)\le Dn+o(n)$,
which is equivalent to
\begin{equation*}
  \Limsup_{n \to \infty}\frac{H(\alpha_n)}{n}\le D.
\end{equation*}
\qed
\end{definition}


In the case of $D=1$,
the weak Chaitin $D$-randomness results in the weak Chaitin randomness.
For any $D\in[0,1]$ and any $\alpha\in\XI$,
if $\alpha$ is weakly Chaitin $D$-random and $D$-compressible,
then
\begin{equation}\label{compression-rate}
  \lim_{n\to \infty} \frac{H(\alpha_n)}{n} = D.
\end{equation}
Hereafter the left-hand side of \eqref{compression-rate}
is referred to as the \textit{compression rate} of
an infinite binary string $\alpha$ in general.
Note, however, that \eqref{compression-rate}
does not necessarily imply that $\alpha$ is weakly Chaitin $D$-random.

In the works \cite{T99,T02},
we generalized Chaitin's halting probability $\Omega$ to $\Omega^D$ by
\eqref{intro-omegad} for any real number $D>0$.
Thus,
$\Omega=\Omega^1$.
If $0<D\le 1$, then $\Omega^D$ converges and $0<\Omega^D<1$,
since $\Omega^D\le \Omega<1$.

\begin{theorem}[Tadaki \cite{T99,T02}]\label{pomgd}
Let $D\in\R$.
\begin{enumerate}
  \item If $0<D\le 1$ and $D$ is computable,
    then $\Omega^D$ is weakly Chaitin $D$-random and
    $D$-compressible.
  \item If $1<D$, then $\Omega^D$ diverges to $\infty$.\qed
\end{enumerate}
\end{theorem}



\begin{definition}[Chaitin $D$-randomness, Tadaki \cite{T99,T02}]
  Let $D\in\R$ with $D\ge 0$,
  and let $\alpha \in\XI$.
  We say that $\alpha$ is \textit{Chaitin $D$-random} if
  $\lim_{n\to \infty} H(\alpha_n)-Dn = \infty$.
  \qed
\end{definition}

In the case of $D=1$,
the Chaitin $D$-randomness results in the Chaitin randomness.
Obviously,
for any $D\in[0,1]$ and any $\alpha\in\XI$,
if $\alpha$ is Chaitin $D$-random,
then $\alpha$ is weakly Chaitin $D$-random.
However,
in 2005 Reimann and Stephan \cite{RS05} showed that,
in the case of $D<1$,
the converse does not necessarily hold.
This contrasts with the equivalence between
the weakly Chaitin randomness and the Chaitin randomness,
each of which corresponds to the case of $D=1$.
In the next section,
for any computable real number $D$ with $0<D<1$,
we give an instance of a real number
which is Chaitin $D$-random and $D$-compressible.

\section{Chaitin $D$-randomness and divergence}
\label{Chaitin D-randomness}

For each real numbers $Q>0$ and $D>0$,
we define $W(Q,D)$ by
\begin{equation*}
  W(Q,D)=
  \sum_{p\in\Dom U}\abs{p}^{Q} 2^{-\frac{\abs{p}}{D}}.
\end{equation*}
As the first
result of this paper,
we show the following theorem.

\begin{theorem}\label{ctr}
Let $Q$ and $D$ be positive real numbers.
\begin{enumerate}
  \item If $Q$ and $D$ are computable and $0<D<1$,
    then $W(Q,D)$ converges to a left-computable real number
    which is Chaitin $D$-random and $D$-compressible.
  \item If $1\le D$, then $W(Q,D)$ diverges to $\infty$.\qed
\end{enumerate}
\end{theorem}

The techniques used in the proofs of
Theorem~\ref{pomgd} (see \cite{T02}) and Theorem~\ref{ctr}
are frequently used throughout
the rest of
this paper as basic tools.
We see that
the weak Chaitin $D$-randomness in Theorem~\ref{pomgd}
is replaced by the Chaitin $D$-randomness
in Theorem~\ref{ctr}
in exchange for the divergence at $D=1$.
In order to derive this divergence
we make use of Theorem~\ref{numerator} (i) below.
We prove Theorem~\ref{numerator} in a more general form,
and show that
the \textit{Shannon entropy} $-\sum_{s\in\X} m(s)\log_2 m(s)$ of
an arbitrary universal probability $m$ diverges to $\infty$.
We say that
a
function $f\colon\N^+\to[0,\infty)$ is \textit{lower-computable}
if there exists a total recursive function $a\colon\N^+\times \N^+\to\Q$
such that, for each $n\in\N^+$,
$\lim_{k\to\infty} a(k,n)=f(n)$ and $\forall\,k\in\N^+\;\,a(k,n)\le f(n)$.

\begin{theorem}\label{numerator}
Let $A$ be an infinite r.e.~subset of $\X$
and let $f\colon\N^+\to[0,\infty)$ be a lower-computable function such that
$\lim_{n\to\infty}f(n)=\infty$.
Then the following hold.
\begin{enumerate}
  \item $\sum_{U(p)\in A}f(\abs{p})2^{-\abs{p}}$ diverges to $\infty$.
  \item If there exists $l_0\in\N^+$ such that
    $f(l)2^{-l}$ is a nonincreasing function of $l$ for all $l\ge l_0$,
    then $\sum_{s\in A} f(H(s))2^{-H(s)}$ diverges to $\infty$.
\end{enumerate}
\end{theorem}

\begin{proof}
(i)
Contrarily, assume that $\sum_{U(p)\in A}f(\abs{p})2^{-\abs{p}}$ converges.
Then, there exists $d\in\N^+$ such that
$\sum_{U(p)\in A}f(\abs{p})2^{-\abs{p}}\le d$.
We define the function $r\colon \X\to [0,\infty)$ by
\begin{equation*}
  r(s)=\frac{1}{d}\sum_{U(p)=s}f(\abs{p})2^{-\abs{p}}
\end{equation*}
if $s\in A$; $r(s)=0$ otherwise.
Then we see that $\sum_{s\in\X}r(s)\le 1$ and
therefore $r$ is a lower-computable semi-measure.
Since $P(s)$ is a universal probability,
there exists $c\in\N^+$ such that $r(s)\le cP(s)$ for all $s\in\X$.
Hence we have
\begin{equation}\label{fp2p}
  \sum_{U(p)=s}(cd-f(\abs{p}))2^{-\abs{p}}\ge 0
\end{equation}
for all $s\in A$.
On the other hand,
since $A$ is an infinite set and $\lim_{n\to\infty}f(n)=\infty$,
there is $s_0\in A$ such that $f(\abs{p})>cd$ for all $p$ with $U(p)=s_0$.
Therefore we have $\sum_{U(p)=s_0}(cd-f(\abs{p}))2^{-\abs{p}}< 0$.
However, this contradicts \eqref{fp2p},
and the proof of (i) is completed.

(ii)
We first note that there is $n_0\in\N$ such that
$H(s)\ge l_0$ for all $s$ with $\abs{s}\ge n_0$.
Now,
let us assume contrarily that $\sum_{s\in A} f(H(s))2^{-H(s)}$ converges.
Then, there exists $d\in\N^+$ such that
$\sum_{s\in A} f(H(s))2^{-H(s)}\le d$.
We define the function $r\colon \X\to [0,\infty)$ by
\begin{equation*}
  r(s)=\frac{1}{d}f(H(s))2^{-H(s)}
\end{equation*}
if $s\in A$ and $\abs{s}\ge n_0$; $r(s)=0$ otherwise.
Then we see that $\sum_{s\in\X}r(s)\le 1$ and
therefore $r$ is a lower-computable semi-measure.
Since $2^{-H(s)}$ is a universal probability by Theorem \ref{eup},
there exists $c\in\N^+$ such that $r(s)\le c2^{-H(s)}$ for all $s\in\X$.
Hence, if $s\in A$ and $\abs{s}\ge n_0$,
then $cd\ge f(H(s))$.
On the other hand,
since $A$ is an infinite set and $\lim_{n\to\infty}f(n)=\infty$,
there is $s_0\in A$ such that $\abs{s_0}\ge n_0$ and $f(H(s_0))>cd$.
Thus, we have a contradiction, and the proof of (ii) is completed.
\end{proof}

\begin{corollary}\label{ShannonA}
If $m$ is a universal probability and $A$ is an infinite r.e.~subset of $\X$,
then $-\sum_{s\in A} m(s)\log_2 m(s)$ diverges to $\infty$.
\end{corollary}

\begin{proof}
We first note that
there is a real number $x_0>0$ such that
the function $x2^{-x}$ of a real number $x$ is decreasing for $x\ge x_0$.
For this $x_0$,
there is $n_0\in\N$ such that
$-\log_2 m(s)\ge x_0$ for all $s$ with $\abs{s}\ge n_0$.
On the other hand,
by \eqref{eq: K_m},
there is $c\in\N$ such that $-\log_2 m(s)\le H(s)+c$ for all $s\in\X$.
Thus, we see that
\begin{align*}
  -\sum_{s\in A\text{ \& }\abs{s}\ge n_0} m(s)\log_2 m(s)
  &\ge
  \sum_{s\in A\text{ \& }\abs{s}\ge n_0}(H(s)+c)2^{-H(s)-c} \\
  &=
  2^{-c}\sum_{s\in A\text{ \& }\abs{s}\ge n_0}H(s)2^{-H(s)}
  +
  c2^{-c}\sum_{s\in A\text{ \& }\abs{s}\ge n_0}2^{-H(s)}.
\end{align*}
By Theorem \ref{numerator} (ii),
$\sum_{s\in A}H(s)2^{-H(s)}$ diverges to $\infty$.
Hence,
we see, by the inequality above,
that $-\sum_{s\in A} m(s)\log_2 m(s)$ also diverges to $\infty$.
\end{proof}

By Corollary \ref{ShannonA},
we see that
the Shannon entropy
of an arbitrary universal probability diverges to $\infty$.

The proof of Theorem~\ref{ctr} is given as follows.

\begin{proof}[The proof of Theorem~\ref{ctr}]
Let $p_1,p_2,p_3,\dots$ be a recursive enumeration of the r.e.~set $\Dom U$.
Then,
for every $D>0$,
$W(Q,D)=\lim_{m\to\infty} \widetilde{W}_m(Q,D)$,
where
\begin{equation*}
  \widetilde{W}_m(Q,D)
  =\sum_{i=1}^{m} \abs{p_i}^{Q}2^{-\frac{\abs{p_i}}{D}}.
\end{equation*}

(i) First we show that $W(Q,D)$ converges
to a left-computable real number.
Since $D<1$,
there is $l_0\in\N^+$ such that
\begin{equation*}
  \frac{1}{D}-\frac{Q\log_2 l}{l}\ge 1
\end{equation*}
for all $l\ge l_0$.
Then there is $m_0\in\N^+$ such that $\abs{p_i}\ge l_0$ for all $i>m_0$.
Thus, we see that,
for each $i>m_0$,
\begin{equation*}
  \abs{p_i}^{Q}2^{-\frac{\abs{p_i}}{D}}
  =2^{-(\frac{1}{D}-\frac{Q\log_2 \abs{p_i}}{\abs{p_i}})\abs{p_i}}
  \le 2^{-\abs{p_i}}.
\end{equation*}
Hence, for each $m>m_0$,
\begin{equation*}
  \widetilde{W}_{m}(Q,D)-\widetilde{W}_{m_0}(Q,D)
  =\sum_{i=m_0+1}^{m} \abs{p_i}^{Q}2^{-\frac{\abs{p_i}}{D}}
  \le \sum_{i=m_0}^{m} 2^{-\abs{p_i}}<\Omega.
\end{equation*}
Thus,
since $\{\widetilde{W}_{m}(Q,D)\}_m$ is
an increasing sequence of real numbers,
it
converges to a real number $W(Q,D)$
as $m\to\infty$.
Moreover,
since $Q$ and $D$ are computable,
$W(Q,D)$ is shown to be left-computable.

We then show that $W(Q,D)$ is Chaitin $D$-random.
Let $\alpha$ be the infinite binary string such that $0.\alpha$ is
the base-two expansion of $W(Q,D)\bmod' 1$ with infinitely many ones.
Then,
since $Q$ and $D$ are computable real numbers
and $\lceil W(Q,D) \rceil -1 + 0.\alpha_n<W(Q,D)$ for all $n\in\N^+$,
there exists a partial recursive function $\xi\colon\X\to \N^+$ such that,
for all $n\in\N^+$,
\begin{equation*}
  \lceil W(Q,D) \rceil -1 + 0.\alpha_n
  < \widetilde{W}_{\xi(\alpha_n)}(Q,D).
\end{equation*}
It is then easy to see that
\begin{equation*}
  W(Q,D)-\widetilde{W}_{\xi(\alpha_n)}(Q,D) < 2^{-n}.
\end{equation*}
It follows that,
for all $i>\xi(\alpha_n)$,
$\abs{p_i}^Q 2^{-\frac{\abs{p_{i}}}{D}}<2^{-n}$
and therefore $QD\log_2\abs{p_i}<\abs{p_i}-Dn$.
Thus, given $\alpha_n$,
by calculating the set
$\left\{
  \>U(p_i)\bigm|i\le \xi(\alpha_n)\;
\right\}$
and picking any
one
finite binary string which is not in this set,
one can obtain $s\in\X$ such that
$QD\log_2H(s)<H(s)-Dn$.

Hence, there exists a partial recursive function $\Psi\colon\X\to\X$
such that
\begin{equation*}
  QD\log_2H(\Psi(\alpha_n))<H(\Psi(\alpha_n))-Dn.
\end{equation*}
Applying this inequality to itself, we have
$QD\log_2n<H(\Psi(\alpha_n))-Dn+O(1)$.
On the other hand,
using \eqref{eq: k} there is a natural number $c_\Psi$ such that
$H(\Psi(\alpha_n))<H(\alpha_n)+c_\Psi$.
Therefore, we have
\begin{equation*}
  QD\log_2n<H(\alpha_n)-Dn+O(1).
\end{equation*}
Hence, $\alpha$ is Chaitin $T$-random.
It follows that $\alpha$ has infinitely many zeros,
which implies that
$W(Q,D)\bmod 1=W(Q,D)\bmod' 1=0.\alpha$ and therefore $(W(Q,D))_n=\alpha_n$.
Thus, $W(Q,D)$ is Chaitin $D$-random.

Next,
we show that $W(Q,D)$ is $D$-compressible.
Since $Q$ and $D$ are computable real numbers,
there exists a total recursive function
$g\colon\N^+\times\N^+\to\Z$ such that,
for all $m,n\in\N^+$,
\begin{equation}\label{eq: str-dfrw}
  \abs{\;
    \widetilde{W}_{m}(Q,D)
    -\lfloor W(Q,D) \rfloor
    -2^{-n}g(m,n)
  \;} <
  2^{-n}.
\end{equation}
Let $d$ be any computable real number with $D<d<1$.
Then,
the limit value $W(Q,d)$ exists since $d<1$.
Let $\beta$ be the infinite binary string such that
$0.\beta$ is the base-two expansion of $W(Q,d)\bmod' 1$
with infinitely many ones.

Given $n$ and $\beta_{\lceil Dn/d\rceil}$
(i.e., the first $\lceil Dn/d\rceil$ bits of $\beta$),
one can find $m_0\in\N^+$ such that
\begin{equation*}
  \lceil W(Q,d) \rceil-1+0.\beta_{\lceil Dn/d\rceil}
  <\widetilde{W}_{m_0}(Q,d).
\end{equation*}
This is possible since
$\lceil W(Q,d) \rceil-1+0.\beta_{\lceil Dn/d\rceil}<W(Q,d)$
and $\lim_{m\to\infty}\widetilde{W}_{m}(Q,d)=W(Q,d)$.
It is then easy to see that
\begin{equation*}
  \sum_{i=m_0+1}^\infty \abs{p_{i}}^Q 2^{-\frac{\abs{p_{i}}}{d}} < 2^{-Dn/d}.
\end{equation*}
Raising both sides of this inequality to the power $d/D$ and
using the inequality $a^c+b^c\le (a+b)^c$
for real numbers $a,b>0$ and $c\ge 1$,
\begin{equation*}
  2^{-n}
  >
  \sum_{i=m_0+1}^\infty \abs{p_i}^{Qd/D} 2^{-\frac{\abs{p_{i}}}{D}}
  >
  \sum_{i=m_0+1}^\infty \abs{p_{i}}^Q 2^{-\frac{\abs{p_{i}}}{D}}.
\end{equation*}
It follows that
\begin{equation}\label{eq: ods-dfrw}
  \abs{\,
    W(Q,D)-\widetilde{W}_{m_0}(Q,D)
  \,}
  <2^{-n}.
\end{equation}
From \eqref{eq: str-dfrw}, \eqref{eq: ods-dfrw}, and
\begin{equation*}
  \abs{\,
    \lfloor W(Q,D) \rfloor + 0.(W(Q,D))_n-W(Q,D)
  \,}
  <2^{-n},
\end{equation*}
it is shown that
\begin{equation*}
  \abs{\;(W(Q,D))_n - g(m_0,n)\;}<3
\end{equation*}
and therefore
\begin{equation*}
  (W(Q,D))_n
  =g(m_0,n),\,g(m_0,n)\pm 1,\,g(m_0,n)\pm 2,
\end{equation*}
where $(W(Q,D))_n$ is regarded as a dyadic integer.
Thus, there are still $5$ possibilities of $(W(Q,D))_n$,
so that one needs only $3$ bits more in order to determine $(W(Q,D))_n$.

Thus, there exists a partial recursive function
$\Phi\colon \N^+\times\X\times\X\to\X$ such that
\begin{equation*}
  \forall\,n\in\N^+\quad\exists\,s\in\X\quad
  \abs{s}=3\;\;\&\;\;
  \Phi(n,\beta_{\lceil Dn/d\rceil},s)=(W(Q,D))_n.
\end{equation*}
It follows from \eqref{eq: fas} that
$H((W(Q,D))_n)
\le |\beta_{\lceil Dn/d\rceil}|+o(n)
\le Dn/d+o(n)$,
which implies that $W(Q,D)$ is $D/d$-compressible.
Since $d$ is any computable real number with $D<d<1$,
it follows that $W(Q,D)$ is $D$-compressible.

(ii)
We choose any one computable real number $Q'$ with $Q\ge Q'>0$.
Then,
using Theorem \ref{numerator} (i),
we can show that
$W(Q',1)=\sum_{p\in\Dom U}\abs{p}^{Q'} 2^{-\abs{p}}$
diverges to $\infty$.
Thus,
since $\widetilde{W}_{m}(Q,D)\ge\widetilde{W}_{m}(Q',1)$,
we see that $W(Q,D)$ also diverges to $\infty$,
and the proof is completed.
\end{proof}

\section{Temperature as a compression rate}
\label{tcr}

In this section
we introduce
the notion of thermodynamic quantities
such as partition function, free energy, energy, entropy, and specific heat,
into algorithmic information theory
by performing Replacements~\ref{CS07}
for the corresponding thermodynamic quantities
in statistical mechanics.%
\footnote{
For the thermodynamic quantities in statistical mechanics,
see Chapter 16 of \cite{C85} and Chapter 2 of \cite{TKS92}.
To be precise,
the partition function is not a thermodynamic quantity
but a statistical mechanical quantity.
}
We investigate their convergence and the degree of randomness.
For that purpose,
we first choose any one enumeration
$q_1,q_2,q_3, \dotsc$ of the countably infinite set $\Dom U$
as the standard one for use throughout
this section.%
\footnote{
The enumeration $\{q_i\}$ is quite arbitrary and therefore
we do not,
ever,
require $\{q_i\}$ to be a recursive enumeration of $\Dom U$.
}

In statistical mechanics,
the partition function $Z_{\text{sm}}(T)$ at temperature $T$ is given by
\begin{equation}\label{partition_function_sm}
  Z_{\text{sm}}(T)=\sum_{x\in X}e^{-\frac{E_x}{kT}}.
\end{equation}
Motivated by the formula \eqref{partition_function_sm}
and taking into account Replacements~\ref{CS07},
we introduce the notion of partition function into
algorithmic information theory
as follows.

\begin{definition}[partition function]\label{def-partition-function}
For each $n\in\N^+$ and each real number $T>0$,
we define $Z_n(T)$ by
\begin{equation*}
  Z_n(T)=\sum_{i=1}^n 2^{-\frac{\abs{q_i}}{T}}.
\end{equation*}
Then, for each $T>0$,
the \textit{partition function} $Z(T)$ is defined by
$Z(T)=\lim_{n\to\infty} Z_n(T)$.
\qed
\end{definition}

Since $Z(T)=\Omega^T$,
we restate Theorem \ref{pomgd} as in the following form.

\begin{theorem}[Tadaki \cite{T99,T02}]\label{crpf}
Let $T\in\R$.
\begin{enumerate}
  \item If $0<T\le 1$ and $T$ is computable,
    then $Z(T)$ converges to a left-computable real number which is
    weakly Chaitin $T$-random and $T$-compressible.
  \item If $1<T$, then $Z(T)$ diverges to $\infty$.\qed
\end{enumerate}
\end{theorem}

In statistical mechanics,
the free energy $F_{\text{sm}}(T)$ at temperature $T$ is given by
\begin{equation}\label{free_energy_sm}
  F_{\text{sm}}(T)=-kT\ln Z_{\text{sm}}(T),
\end{equation}
where $Z_{\text{sm}}(T)$ is given by \eqref{partition_function_sm}.
Motivated by the formula \eqref{free_energy_sm}
and taking into account Replacements~\ref{CS07},
we introduce the notion of free energy into
algorithmic information theory
as follows.

\begin{definition}[free energy]\label{def-free-energy}
For each $n\in\N^+$ and each real number $T>0$,
we define $F_n(T)$ by
\begin{equation*}
  F_n(T)=-T\log_2 Z_n(T).
\end{equation*}
Then, for each $T>0$,
the \textit{free energy} $F(T)$ is defined by
$F(T)=\lim_{n\to\infty} F_n(T)$.
\qed
\end{definition}


\begin{theorem}\label{free-energy}
Let $T\in\R$.
\begin{enumerate}
  \item If $0<T\le 1$ and $T$ is computable,
    then $F(T)$ converges to a right-computable real number which is
    weakly Chaitin $T$-random and $T$-compressible.
  \item If $1<T$, then $F(T)$ diverges to $-\infty$.
\end{enumerate}
\end{theorem}


\begin{proof}
(i)
Since $Z(T)$ converges
by Theorem \ref{crpf} (i) and $Z(T)>0$,
$F(T)$ also converges and
\begin{equation*}
  F(T)=-T\log_2 Z(T).
\end{equation*}
Note that
$T$ is a right-computable real number and $-\log_2 Z(T)>0$.
Since $Z(T)$ is a left-computable real number by Theorem \ref{crpf} (i),
$F(T)$ is a right-computable real number.

We show that $F(T)$ is weakly Chaitin $T$-random.
By the mean value theorem,
there exists $c\in\N^+$ such that,
for any $A,B\in\R$,
if $A\ge F(T)$ and $B\ge 1/T$ then
\begin{equation}\label{mvt_ft}
  0\le Z(T)-2^{-AB}\le 2^c\max\{A-F(T),B-1/T\}.
\end{equation}
Since $F(T)$ is a right-computable real number,
there exists a total recursive function $f\colon\N^+\to\Q$ such that
$F(T)\le f(m)$ for all $m\in\N^+$ and $\lim_{m\to\infty} f(m)=F(T)$.
Since $T$ is a computable real number,
there exists a total recursive function $g\colon\N^+\to\Q$ such that
$0\le g(n)-1/T<2^{-n}$ for all $n\in\N^+$.

Given $(F(T))_n$,
one can find $m_0\in\N^+$ such that
\begin{equation*}
  f(m_0)<\lfloor F(T) \rfloor +0.(F(T))_{n}+2^{-n}.
\end{equation*}
This is possible because
$F(T)<\lfloor F(T) \rfloor +0.(F(T))_{n}+2^{-n}$.
It follows that
$0\le f(m_0)-F(T)<2^{-n}$.
Therefore, by \eqref{mvt_ft} it is shown that
$0\le Z(T)-2^{-f(m_0)g(n)}<2^{c-n}$.
Let $l_n$ be the first $n$ bits of the base-two expansion of
$2^{-f(m_0)g(n)}$
with infinitely many zeros.
It follows that
\begin{equation*}
  0\le 0.(Z(T))_{n}-0.l_{n}
  <Z(T)-2^{-f(m_0)g(n)}+2^{-n}
  <(2^c+1)2^{-n}.
\end{equation*}
Hence
\begin{equation*}
  (Z(T))_n=l_n,\,l_n+1,\,l_n+2,\,\dots,\,l_n+(2^c+1),
\end{equation*}
where $(Z(T))_n$ and $l_n$ are regarded as a dyadic integer.
Therefore, there are still $2^c+2$ possibilities of $(Z(T))_n$,
so that one needs only $c+1$ bits more in order to determine $(Z(T))_n$.

Thus, there exists a partial recursive function
$\Phi\colon \X\times\X\to\X$ such that
\begin{equation*}
  \forall\,n\in\N^+\quad\exists\,s\in\X\quad
  \abs{s}=c+1\;\;\&\;\;
  \Phi((F(T))_n,s)=(Z(T))_n.
\end{equation*}
It follows that there exists $c_{\Phi}\in\N^+$ such that,
for all $n\in\N^+$,
\begin{equation*}
  H((Z(T))_n) \le H((F(T))_n)+c_{\Phi}.
\end{equation*}
Hence, $F(T)$ is weakly Chaitin $T$-random by Theorem \ref{crpf} (i).

Next,
we show that $F(T)$ is $T$-compressible.
Since $T$ is a computable real number,
there exists a total recursive function $a\colon\X\times\N^+\to\Z$
such that,
for all $s\in\X$ and all $n\in\N^+$, if $0.s>0$ then
\begin{equation}\label{a-tcfe}
  \abs{\,
    -T\log_2 0.s
    -\lfloor F(T) \rfloor
    -2^{-n}a(s,n)
  \,}
  <
  2^{-n}.
\end{equation}
By the mean value theorem,
it is also shown that there is $d\in\N^+$ such that,
for all $n\in\N^+$, if $0.(Z(T))_n>0$ then
\begin{equation}\label{Kotake-tcfe}
  \abs{\,
    -T\log_2 0.(Z(T))_n-F(T)
  \,}
  <
  2^{d-n}.
\end{equation}
From \eqref{a-tcfe}, \eqref{Kotake-tcfe}, and
\begin{equation*}
  \abs{\,
    \lfloor F(T) \rfloor + 0.(F(T))_n-F(T)
  \,}
  <2^{-n},
\end{equation*}
it is shown that,
for all $n\in\N^+$, if $0.(Z(T))_n>$ then
\begin{equation*}
  \abs{\;(F(T))_n - a((Z(T))_n,n)\;}<2^d+2
\end{equation*}
and therefore
\begin{equation*}
\begin{split}
  (F(T))_n
  &=a((Z(T))_n,n),\,a((Z(T))_n,n)\pm 1,\,a((Z(T))_n,n)\pm 2,\,\dots,\, \\
  &\hspace*{4.7mm}
  a((Z(T))_n,n)\pm (2^d+1),
\end{split}
\end{equation*}
where $(F(T))_n$ is regarded as a dyadic integer.
Therefore, there are still $2^{d+1}+3$ possibilities of $(F(T))_n$,
so that one needs only $d+2$ bits more in order to determine $(F(T))_n$.

Thus, there exists a partial recursive function
$\Psi\colon \X\times\X\to\X$ such that
\begin{equation*}
  \forall\,n\in\N^+\quad\exists\,s\in\X\quad
  \abs{s}=d+2\;\;\&\;\;
  \Psi((Z(T))_n,s)=(F(T))_n.
\end{equation*}
It follows that there exists $c_{\Psi}\in\N^+$ such that,
for all $n\in\N^+$,
\begin{equation*}
  H((F(T))_n) \le H((Z(T))_n)+c_{\Psi}.
\end{equation*}
Since $Z(T)$ is $T$-compressible by Theorem \ref{crpf} (i),
$F(T)$ is also $T$-compressible.

(ii)
In the case of $T>1$,
since $\lim_{n\to\infty}Z_n(T)=\infty$ by Theorem \ref{crpf} (ii),
we see that
$F_n(T)$ diverges to $-\infty$ as $n\to\infty$.
\end{proof}

In statistical mechanics,
the energy $E_{\text{sm}}(T)$ at temperature $T$ is given by
\begin{equation}\label{energy_sm}
  E_{\text{sm}}(T)
  =\frac{1}{Z_{\text{sm}}(T)}\sum_{x\in X}E_xe^{-\frac{E_x}{kT}},
\end{equation}
where $Z_{\text{sm}}(T)$ is given by \eqref{partition_function_sm}.
Motivated by the formula \eqref{energy_sm}
and taking into account Replacements~\ref{CS07},
we introduce the notion of energy
into algorithmic information theory
as follows.

\begin{definition}[energy]\label{def-energy}
For each $n\in\N^+$ and each real number $T>0$,
we define $E_n(T)$ by
\begin{equation*}
  E_n(T)=\frac{1}{Z_n(T)}\sum_{i=1}^n \abs{q_i}2^{-\frac{\abs{q_i}}{T}}.
\end{equation*}
Then, for each $T>0$,
the \textit{energy} $E(T)$ is defined by
$E(T)=\lim_{n\to\infty} E_n(T)$.
\qed
\end{definition}

\begin{theorem}\label{energy}
Let $T\in\R$.
\begin{enumerate}
  \item If $0<T<1$ and $T$ is computable,
    then $E(T)$ converges to a left-computable real number which is
    Chaitin $T$-random and $T$-compressible.
  \item If $1\le T$, then $E(T)$ diverges to $\infty$.
\end{enumerate}
\end{theorem}

\begin{proof}
(i) First we show that $E(T)$ converges.
By Theorem~\ref{crpf} (i),
the denominator $Z_n(T)$ of $E_n(T)$ converges
to the real number $Z(T)>0$
as $n\to\infty$.
On the other hand,
by Theorem~\ref{ctr} (i),
the numerator $\sum_{i=1}^n \abs{q_i}2^{-\frac{\abs{q_i}}{T}}$ of $E_n(T)$
converges to the real number $W(1,T)$
as $n\to\infty$.
Thus,
$E_n(T)$ converges to the real number $W(1,T)/Z(T)$ as $n\to\infty$.

Next, we show that $E(T)$ is a left-computable real number.
Let $p_1,p_2,p_3,\dots$ be a recursive enumeration of the r.e.~set $\Dom U$.
For each $m\in\N^+$,
we define $\widetilde{W}_m(T)$ and $\widetilde{Z}_m(T)$ by
\begin{equation*}
  \widetilde{W}_m(T)=\sum_{i=1}^{m} \abs{p_i}2^{-\frac{\abs{p_i}}{T}}
  \quad\text{ and }\quad
  \widetilde{Z}_m(T)=\sum_{i=1}^{m} 2^{-\frac{\abs{p_i}}{T}},
\end{equation*}
and then define $\widetilde{E}_m(T)$ by
$\widetilde{E}_m(T)=\widetilde{W}_m(T)/\widetilde{Z}_m(T)$.
Since the numerator and the denominator of $E_n(T)$ are positive term series
which converge as $n\to\infty$,
we see that
$W(1,T)=\lim_{m\to\infty}\widetilde{W}_m(T)$,
$Z(T)=\lim_{m\to\infty}\widetilde{Z}_m(T)$, and
$E(T)=\lim_{m\to\infty} \widetilde{E}_m(T)$.
%
We then see that
\begin{equation*}
  \widetilde{E}_{m+1}(T)-\widetilde{E}_{m}(T)
  =
  \frac{\widetilde{Z}_{m}(T)\abs{p_{m+1}}-\widetilde{W}_{m}(T)}
  {\widetilde{Z}_{m+1}(T)\widetilde{Z}_{m}(T)}
  2^{-\frac{\abs{p_{m+1}}}{T}}.
\end{equation*}
Since $\widetilde{W}_m(T)$ and $\widetilde{Z}_m(T)$ converge
as $m\to\infty$
and $\lim_{m\to\infty}\abs{p_m}=\infty$,
there exist $a\in\N^+$ and $m_1\in\N^+$
such that, for any $m\ge m_1$,
\begin{equation}\label{dfrncL}
  \widetilde{E}_{m+1}(T)-\widetilde{E}_{m}(T)
  >\abs{p_{m+1}}2^{-\frac{\abs{p_{m+1}}}{T}-a}.
\end{equation}
In particular,
$\widetilde{E}_{m}(T)$ is
an increasing function of $m$ for all $m\ge m_1$ by the above inequality.
Thus, since $T$ is a computable real number,
$E(T)$ is shown to be a left-computable real number.

We then show that $E(T)$ is Chaitin $T$-random.
Let $\alpha$ be the infinite binary string such that $0.\alpha$ is
the base-two expansion of $E(T)\bmod' 1$ with infinitely many ones.
Then,
since $T$ is a computable real number and
$\lceil E(T) \rceil -1 + 0.\alpha_n < E(T)$ for all $n\in\N^+$,
there exists a partial recursive function $\xi\colon\X\to \N^+$ such that,
for all $n\in\N^+$,
$\xi(\alpha_n)\ge m_1$ and
\begin{equation*}
  \lceil E(T) \rceil -1 + 0.\alpha_n < \widetilde{E}_{\xi(\alpha_n)}(T).
\end{equation*}
It is then easy to see that
$E(T)-\widetilde{E}_{\xi(\alpha_n)}(T) < 2^{-n}$.
It follows from \eqref{dfrncL} that,
for all $i>\xi(\alpha_n)$,
$\abs{p_i}2^{-\frac{\abs{p_{i}}}{T}-a}<2^{-n}$
and therefore $T\log_2\abs{p_i}-Ta<\abs{p_i}-Tn$.
Thus, given $\alpha_n$,
by calculating the set
$\left\{
  \>U(p_i)\bigm|i\le \xi(\alpha_n)\;
\right\}$
and picking any
one
finite binary string which is not in this set,
one can obtain $s\in\X$ such that
$T\log_2H(s)-Ta<H(s)-Tn$.

Hence, there exists a partial recursive function $\Psi\colon\X\to\X$
such that
\begin{equation*}
  T\log_2H(\Psi(\alpha_n))-Ta<H(\Psi(\alpha_n))-Tn.
\end{equation*}
Applying this inequality to itself, we have
$T\log_2n<H(\Psi(\alpha_n))-Tn+O(1)$.
On the other hand,
using \eqref{eq: k} there is a natural number $c_\Psi$ such that
$H(\Psi(\alpha_n))<H(\alpha_n)+c_\Psi$.
Therefore, we have
\begin{equation*}
  T\log_2n<H(\alpha_n)-Tn+O(1).
\end{equation*}
Hence, $\alpha$ is Chaitin $T$-random.
It follows that $\alpha$ has infinitely many zeros,
which implies that
$E(T)\bmod 1=E(T)\bmod' 1=0.\alpha$ and therefore $(E(T))_n=\alpha_n$.
Thus, $E(T)$ is Chaitin $T$-random.

Next,
we show that $E(T)$ is $T$-compressible.
Since $T$ is a computable real number,
there exists a total recursive function
$g\colon\N^+\times\N^+\to\Z$ such that,
for all $m,n\in\N^+$,
\begin{equation}\label{eq: str-dfr}
  \abs{\;
    \widetilde{E}_{m}(T)
    -\lfloor E(T) \rfloor
    -2^{-n}g(m,n)
  \;} <
  2^{-n}.
\end{equation}
It is also shown that there is $c\in\N^+$ such that,
for all $m\in\N^+$,
\begin{equation}\label{pfKoatke}
  \abs{\;
    E(T)-\widetilde{E}_{m}(T)
  \;}
  <
  2^c
  \max
  \left\{
  \abs{\,
    W(1,T)-\widetilde{W}_{m}(T)
  \,},
  \abs{\,
    Z(T)-\widetilde{Z}_{m}(T)
  \,}
  \right\}.
\end{equation}
Let $t$ be any computable real number with $T<t<1$,
Then,
$W(1,t)=\lim_{m\to\infty}\widetilde{W}_m(t)$,
where
\begin{equation*}
  \widetilde{W}_m(t)
  =\sum_{i=1}^{m} \abs{p_i} 2^{-\frac{\abs{p_i}}{t}}.
\end{equation*}
The limit value $W(1,t)$ exists since $t<1$.
Let $\beta$ be the infinite binary string such that
$0.\beta$ is the base-two expansion of $W(1,t)\bmod' 1$
with infinitely many ones.

Given $n$ and $\beta_{\lceil Tn/t\rceil}$
(i.e., the first $\lceil Tn/t\rceil$ bits of $\beta$),
one can find $m_0\in\N^+$ such that
\begin{equation*}
  \lceil W(1,t) \rceil-1+0.\beta_{\lceil Tn/t\rceil}
  <\widetilde{W}_{m_0}(t).
\end{equation*}
This is possible since
$t$ is a computable real number and
$\lceil W(1,t) \rceil-1+0.\beta_{\lceil Tn/t\rceil}<W(1,t)$.
It is then easy to see that
\begin{equation*}
  \sum_{i=m_0+1}^\infty \abs{p_{i}}2^{-\frac{\abs{p_{i}}}{t}} < 2^{-Tn/t}.
\end{equation*}
Raising both sides of this inequality to the power $t/T$ and
using the inequality $a^c+b^c\le (a+b)^c$
for real numbers $a,b>0$ and $c\ge 1$,
\begin{equation*}
  2^{-n}
  >
  \sum_{i=m_0+1}^\infty \abs{p_i}^{t/T} 2^{-\frac{\abs{p_{i}}}{T}}
  >
  \sum_{i=m_0+1}^\infty \abs{p_{i}}2^{-\frac{\abs{p_{i}}}{T}}
\end{equation*}
and therefore
\begin{equation*}
  2^{-n}
  >
  \sum_{i=m_0+1}^\infty 2^{-\frac{\abs{p_{i}}}{T}}.
\end{equation*}
It follows that
\begin{equation}\label{eq: ods-dfr}
  \abs{\,
    W(1,T)-\widetilde{W}_{m_0}(T)
  \,}
  <2^{-n}
  \quad\text{and}\quad
  \abs{\,
    Z(T)-\widetilde{Z}_{m_0}(T)
  \,}
  <2^{-n}.
\end{equation}
From \eqref{eq: str-dfr}, \eqref{pfKoatke}, \eqref{eq: ods-dfr}, and
\begin{equation*}
  \abs{\,
    \lfloor E(T) \rfloor + 0.(E(T))_n-E(T)
  \,}
  <2^{-n},
\end{equation*}
it is shown that
\begin{equation*}
  \abs{\;(E(T))_n - g(m_0,n)\;}<2^c+2
\end{equation*}
and therefore
\begin{equation*}
  (E(T))_n
  =g(m_0,n),\,g(m_0,n)\pm 1,\,g(m_0,n)\pm 2,\,\dots,\,g(m_0,n)\pm (2^c+1),
\end{equation*}
where $(E(T))_n$ is regarded as a dyadic integer.
Thus, there are still $2^{c+1}+3$ possibilities of $(E(T))_n$,
so that one needs only $c+2$ bits more in order to determine $(E(T))_n$.

Thus, there exists a partial recursive function
$\Phi\colon \N^+\times\X\times\X\to\X$ such that
\begin{equation*}
  \forall\,n\in\N^+\quad\exists\,s\in\X\quad
  \abs{s}=c+2\;\;\&\;\;
  \Phi(n,\beta_{\lceil Tn/t\rceil},s)=(E(T))_n.
\end{equation*}
It follows from \eqref{eq: fas} that
$H((E(T))_n)
\le |\beta_{\lceil Tn/t\rceil}|+o(n)
\le Tn/t+o(n)$,
which implies that $E(T)$ is $T/t$-compressible.
Since $t$ is any computable real number with $T<t<1$,
it follows that $E(T)$ is $T$-compressible.

(ii)
In the case of $T=1$,
by Theorem~\ref{ctr} (ii),
the numerator $W(1,1)=\sum_{p\in\Dom U}\abs{p}2^{-\abs{p}}$
of $E(1)$ diverges to $\infty$.
On the other hand, the denominator $Z(1)$ of $E(1)$ converges.
Thus, $E(1)$ diverges to $\infty$.

The case of $T>1$ is treated as follows.
We note that $\lim_{n\to\infty}\abs{q_n}=\infty$.
Given $M>0$,
there is $n_0\in\N^+$ such that
$\abs{q_i}\ge 2M$ for all $i>n_0$.
Since $\lim_{n\to\infty}Z_n(T)=\infty$
by Theorem~\ref{crpf} (ii),
there is $n_1\in\N^+$ such that
\begin{equation*}
  \frac{1}{Z_n(T)}
  \sum_{i=1}^{n_0} 2^{-\frac{\abs{q_i}}{T}}\le\frac{1}{2}
\end{equation*}
for all $n>n_1$.
Thus,
for every $n>\max\{n_0,n_1\}$,
\begin{align*}
  E_n(T)
  &=
  \frac{1}{Z_n(T)}
  \sum_{i=1}^{n_0} \abs{q_i}2^{-\frac{\abs{q_i}}{T}}
  +
  \frac{1}{Z_n(T)}
  \sum_{i=n_0+1}^{n} \abs{q_i}2^{-\frac{\abs{q_i}}{T}} \\
  &>
  \frac{2M}{Z_n(T)}
  \sum_{i=n_0+1}^{n} 2^{-\frac{\abs{q_i}}{T}} \\
  &=
  2M
  \left(
    1-
    \frac{1}{Z_n(T)}
    \sum_{i=1}^{n_0} 2^{-\frac{\abs{q_i}}{T}}
  \right) \\
  &\ge 2M\frac{1}{2}=M.
\end{align*}
Hence, $\lim_{n\to\infty} E_n(T)=\infty$,
and the proof is completed.
\end{proof}

In statistical mechanics,
the entropy $S_{\text{sm}}(T)$ at temperature $T$ is given by
\begin{equation}\label{entropy_sm}
  S_{\text{sm}}(T)=\frac{1}{T}E_{\text{sm}}(T)+k\ln Z_{\text{sm}}(T),
\end{equation}
where $Z_{\text{sm}}(T)$ and $E_{\text{sm}}(T)$ are given
by \eqref{partition_function_sm} and \eqref{energy_sm}, respectively.
Motivated by the formula \eqref{entropy_sm}
and taking into account Replacements~\ref{CS07},
we introduce the notion of statistical mechanical entropy into
algorithmic information theory
as follows.

\begin{definition}[statistical mechanical entropy]\label{def-entropy}
For each $n\in\N^+$ and each real number $T>0$,
we define $S_n(T)$ by
\begin{equation*}
  S_n(T)=\frac{1}{T}E_n(T)+\log_2 Z_n(T).
\end{equation*}
Then, for each $T>0$,
the \textit{statistical mechanical entropy} $S(T)$ is defined by
$S(T)=\lim_{n\to\infty} S_n(T)$.
\qed
\end{definition}

\begin{theorem}\label{entropy}
Let $T\in\R$.
\begin{enumerate}
  \item If $0<T<1$ and $T$ is computable,
    then $S(T)$ converges to a left-computable real number which is
    Chaitin $T$-random and $T$-compressible.
  \item If $1\le T$, then $S(T)$ diverges to $\infty$.
\end{enumerate}
\end{theorem}

\begin{proof}
(i)
Since $Z(T)$ and $E(T)$ converge
by Theorem \ref{crpf} (i) and Theorem \ref{energy} (i), respectively,
$S(T)$ also converges and
\begin{equation*}
  S(T)=\frac{1}{T}E(T)+\log_2 Z(T).
\end{equation*}

Since $E(T)$ is a left-computable real number and $E(T)\notin\N$
by Theorem \ref{energy} (i),
there exists a total recursive function $f\colon\N^+\to\Q$ such that
$0< f(m)\le E(T)$ and $\lfloor f(m) \rfloor=\lfloor E(T) \rfloor$
for all $m\in\N^+$ and $\lim_{m\to\infty} f(m)=E(T)$.
Since $T$ is a computable real number,
there exists a total recursive function $g\colon\N^+\to\Q$ such that
$0\le g(m)\le 1/T$ for all $m\in\N^+$ and $\lim_{m\to\infty} g(m)=1/T$.
Since $Z(T)$ is a left-computable real numbers by Theorem \ref{crpf} (i),
there exists a total recursive function $h\colon\N^+\to\Q$ such that
$h(m)\le \log_2 Z(T)$ for all $m\in\N^+$ and
$\lim_{m\to\infty} h(m)=\log_2 Z(T)$.
Hence, $g(m)f(m)+h(m)\le S(T)$ for all $m\in\N^+$ and
$\lim_{m\to\infty}g(m)f(m)+h(m)=S(T)$.
Thus, $S(T)$ is a left-computable real number.

We
then
show that $S(T)$ is weakly Chaitin $T$-random.
Let $\alpha$ be the infinite binary string such that $0.\alpha$ is
the base-two expansion of $S(T)\bmod' 1$ with infinitely many ones.

Given $\alpha_n$,
one can find $m_0\in\N^+$ such that
\begin{equation*}
  \lceil S(T) \rceil-1+0.\alpha_n<g(m_0)f(m_0)+h(m_0).
\end{equation*}
This is possible because
$\lceil S(T) \rceil-1+0.\alpha_n<S(T)$
and $\lim_{m\to\infty}g(m)f(m)+h(m)=S(T)$.
It is shown that
\begin{equation*}
\begin{split}
  2^{-n}
  &>\left(\frac{1}{T}E(T)+\log_2 Z(T)\right)
  -\left(g(m_0)f(m_0)+h(m_0)\right) \\
  &\ge\frac{1}{T}E(T)-g(m_0)f(m_0) \\
  &\ge\frac{1}{T}(E(T)-f(m_0)) \\
  &\ge E(T)-f(m_0).
\end{split}
\end{equation*}
Thus, $0\le E(T)-f(m_0)<2^{-n}$.
Let $l_n$ be the first $n$ bits of the base-two expansion of
the rational number $f(m_0)-\lfloor f(m_0) \rfloor$
with infinitely many zeros.
It follows that
\begin{equation*}
  0\le 0.(E(T))_{n}-0.l_{n}<E(T)-f(m_0)+2^{-n}<2\cdot 2^{-n}.
\end{equation*}
Hence
\begin{equation*}
  (E(T))_n=l_n,\,l_n+1,
\end{equation*}
where $(E(T))_n$ and $l_n$ are regarded as a dyadic integer.
Thus, there are still $2$ possibilities of $(E(T))_n$,
so that one needs only $1$ bit more in order to determine $(E(T))_n$.

Thus, there exists a partial recursive function
$\Phi\colon \X\times\{0,1\}\to\X$ such that
\begin{equation*}
  \forall\,n\in\N^+\quad\exists\,b\in\{0,1\}\quad
  \Phi(\alpha_{n},b)=(E(T))_n.
\end{equation*}
It follows that there exists $c_{\Phi}\in\N^+$ such that,
for all $n\in\N^+$,
\begin{equation*}
  H((E(T))_n)
  \le H(\alpha_n)+c_{\Phi}.
\end{equation*}
Hence, $\alpha$ is Chaitin $T$-random
by Theorem \ref{energy} (i).
It follows that $\alpha$ has infinitely many zeros,
which implies that
$S(T)\bmod 1=S(T)\bmod' 1=0.\alpha$ and therefore $(S(T))_n=\alpha_n$.
Thus, $S(T)$ is also Chaitin $T$-random.

Next,
we show that $S(T)$ is $T$-compressible.
Let $p_1,p_2,p_3,\dots$ be a recursive enumeration of the r.e.~set $\Dom U$.
For each $m\in\N^+$,
we define $\widetilde{W}_m(T)$ and $\widetilde{Z}_m(T)$ by
\begin{equation*}
  \widetilde{W}_m(T)=\sum_{i=1}^{m} \abs{p_i}2^{-\frac{\abs{p_i}}{T}}
  \quad\text{ and }\quad
  \widetilde{Z}_m(T)=\sum_{i=1}^{m} 2^{-\frac{\abs{p_i}}{T}},
\end{equation*}
and then define $\widetilde{S}_m(T)$ by
\begin{equation*}
  \widetilde{S}_m(T)
  =
  \frac{1}{T}\frac{\widetilde{W}_{m}(T)}{\widetilde{Z}_{m}(T)}
  +\log_2\widetilde{Z}_{m}(T).
\end{equation*}
Since, in the definition of $S_n(T)$,
$Z_n(T)$ and
the numerator and the denominator of $E_n(T)$ are positive term series
which converge as $n\to\infty$,
we see that
$W(1,T)=\lim_{m\to\infty}\widetilde{W}_m(T)$,
$Z(T)=\lim_{m\to\infty}\widetilde{Z}_m(T)$, and
$S(T)=\lim_{m\to\infty} \widetilde{S}_m(T)$.

Since $T$ is a computable real number,
there exists a total recursive function $a\colon\N^+\times\N^+\to\Z$ such that,
for all $m,n\in\N^+$,
\begin{equation}\label{a-tcentropy}
  \abs{\;
    \widetilde{S}_m(T)
    -\lfloor S(T) \rfloor
    -2^{-n}a(m,n)
  \;} <
  2^{-n}.
\end{equation}
It is also shown that there is $d\in\N^+$ such that,
for all $m\in\N^+$,
\begin{equation}\label{Kotake-tcentropy}
  \abs{\>
    S(T)-\widetilde{S}_m(T)
  \>}
  <
  2^d
  \max
  \left\{
  \abs{\,
    W(1,T)-\widetilde{W}_{m}(T)
  \,},
  \abs{\,
    Z(T)-\widetilde{Z}_{m}(T)
  \,}
  \right\}.
\end{equation}

Based on the inequalities \eqref{a-tcentropy} and \eqref{Kotake-tcentropy},
in the same manner as the proof of the $T$-compressibility of $E(T)$
in Theorem \ref{energy} (i),
we can show that $S(T)$ is $T$-compressible.

(ii)
In the case of $T\ge 1$,
since $\lim_{n\to\infty}E_n(T)=\infty$ by Theorem \ref{energy} (ii)
and $\log_2 Z_n(T)$ is bounded to the below by Theorem \ref{crpf},
we see that $\lim_{n\to\infty}S_n(T)=\infty$.
\end{proof}

Finally, in statistical mechanics,
the specific heat $C_{\text{sm}}(T)$ at temperature $T$ is given by
\begin{equation}\label{specific_heat_sm}
  C_{\text{sm}}(T)=\frac{d}{dT}E_{\text{sm}}(T),
\end{equation}
where $E_{\text{sm}}(T)$ is given by \eqref{energy_sm}.
Motivated by the formula \eqref{specific_heat_sm},
we introduce the notion of specific heat into
algorithmic information theory
as follows.

\begin{definition}[specific heat]\label{def-specific-heat}
For each $n\in\N^+$ and each real number $T>0$,
we define $C_n(T)$ by
\begin{equation*}
  C_n(T)=E_n'(T),
\end{equation*}
where $E_n'(T)$ is the derived function of $E_n(T)$.
Then, for each $T>0$, the \textit{specific heat} $C(T)$ is defined by
$C(T)=\lim_{n\to\infty} C_n(T)$.
\qed
\end{definition}


\begin{theorem}\label{specific_heat}
Let $T\in\R$.
\begin{enumerate}
  \item If $0<T<1$ and $T$ is computable,
    then $C(T)$ converges to a left-computable real number which is
    Chaitin $T$-random and $T$-compressible,
    and moreover
    $C(T)=E'(T)$
    where $E'(T)$ is the derived function of $E(T)$.
  \item If $T=1$, then $C(T)$ diverges to $\infty$.
\end{enumerate}
\end{theorem}

\begin{proof}
(i) First we show that $C(T)$ converges.
Note that
\begin{equation*}
  C_n(T)
  =
  \frac{\ln 2}{T^2}
  \left\{
    \frac{Y_n(T)}{Z_n(T)}-\left(\frac{W_n(T)}{Z_n(T)}\right)^2
  \right\},
\end{equation*}
where
\begin{equation*}
  Y_n(T)=\sum_{i=1}^{n} \abs{q_i}^22^{-\frac{\abs{q_i}}{T}}
  \quad\text{ and }\quad
  W_n(T)=\sum_{i=1}^{n} \abs{q_i}2^{-\frac{\abs{q_i}}{T}}.
\end{equation*}
By Theorem~\ref{crpf} (i),
$Z_n(T)$ converges to the real number $Z(T)>0$ as $n\to\infty$.
On the other hand,
by Theorem~\ref{ctr} (i),
$Y_n(T)$ and $W_n(T)$ converge to
the real numbers $W(2,T)$ and $W(1,T)$, respectively,
as $n\to\infty$.
Thus,
$C_n(T)$ also converges to a real number $C(T)$ as $n\to\infty$.

Next, we show that $C(T)$ is a left-computable real number.
Let $p_1,p_2,p_3,\dots$ be a recursive enumeration of the r.e.~set $\Dom U$.
For each $m\in\N^+$,
we define $\widetilde{Y}_m(T)$, $\widetilde{W}_m(T)$,
and $\widetilde{Z}_m(T)$ by
\begin{equation*}
  \widetilde{Y}_m(T)=\sum_{i=1}^{m} \abs{p_i}^2 2^{-\frac{\abs{p_i}}{T}},
  \quad
  \widetilde{W}_m(T)=\sum_{i=1}^{m} \abs{p_i}2^{-\frac{\abs{p_i}}{T}},
  \quad
  \text{ and }
  \quad
  \widetilde{Z}_m(T)=\sum_{i=1}^{m} 2^{-\frac{\abs{p_i}}{T}},
\end{equation*}
respectively.
Since $Y_n(T)$, $W_n(T)$, and $Z_n(T)$ are positive term series
which converge as $n\to\infty$,
we see that
$W(2,T)=\lim_{m\to\infty}\widetilde{Y}_m(T)$,
$W(1,T)=\lim_{m\to\infty}\widetilde{W}_m(T)$,
$Z(T)=\lim_{m\to\infty}\widetilde{Z}_m(T)$, and
$C(T)=\lim_{m\to\infty} \widetilde{C}_m(T)$
where
\begin{equation*}
  \widetilde{C}_m(T)
  =
  \frac{\ln 2}{T^2}
  \left\{
    \frac{\widetilde{Y}_m(T)}{\widetilde{Z}_m(T)}
    -\left(\frac{\widetilde{W}_m(T)}{\widetilde{Z}_m(T)}\right)^2
  \right\}.
\end{equation*}
We then see that
$\widetilde{C}_{m+1}(T)-\widetilde{C}_{m}(T)$ is calculated as
\begin{equation}\label{difference-sp}
\begin{split}
  &\frac{\ln 2}{T^2}
  \frac{2^{-\frac{\abs{p_{m+1}}}{T}}}{\widetilde{Z}_{m+1}(T)}
  \Biggl[
    \abs{p_{m+1}}^2
    -
    \left\{
      \frac{\widetilde{W}_{m+1}(T)}{\widetilde{Z}_{m+1}(T)}
      +
      \frac{\widetilde{W}_{m}(T)}{\widetilde{Z}_{m}(T)}
    \right\}
    \abs{p_{m+1}} \\
    &\hspace*{25mm}+
    \left\{
      \frac{\widetilde{W}_{m+1}(T)}{\widetilde{Z}_{m+1}(T)}
      +
      \frac{\widetilde{W}_{m}(T)}{\widetilde{Z}_{m}(T)}
    \right\}
    \frac{\widetilde{W}_{m}(T)}{\widetilde{Z}_{m}(T)}
    -
    \frac{\widetilde{Y}_{m}(T)}{\widetilde{Z}_{m}(T)}
  \Biggr].
\end{split}
\end{equation}
Since $\widetilde{Y}_m(T)$, $\widetilde{W}_m(T)$,
and $\widetilde{Z}_m(T)$ converge as $m\to\infty$
and $\lim_{m\to\infty}\abs{p_m}=\infty$,
there exists $m_1\in\N^+$
such that, for any $m\ge m_1$,
\begin{equation}\label{dfrncC}
  \widetilde{C}_{m+1}(T)-\widetilde{C}_{m}(T)
  >\abs{p_{m+1}}2^{-\frac{\abs{p_{m+1}}}{T}}.
\end{equation}
In particular,
$\widetilde{C}_{m}(T)$ is
an increasing function of $m$ for all $m\ge m_1$ by the above inequality.
Thus, since $T$ is a computable real number,
$C(T)$ is shown to be a left-computable real number.

Based on the computability of $T$ and the inequality \eqref{dfrncC},
in the same manner as the proof of Theorem \ref{energy} (i)
we can show that $C(T)$ is Chaitin $T$-random.

Next,
we show that $C(T)$ is $T$-compressible.
Since $T$ is a computable real number,
there exists a total recursive function
$g\colon\N^+\times\N^+\to\Z$ such that,
for all $m,n\in\N^+$,
\begin{equation}\label{eq: str-dfrC}
  \abs{\;
    \widetilde{C}_{m}(T)
    -\lfloor C(T) \rfloor
    -2^{-n}g(m,n)
  \;} <
  2^{-n}.
\end{equation}
It is also shown that there is $c\in\N^+$ such that,
for all $m\in\N^+$,
\begin{equation}\label{pfKoatkeC}
\begin{split}
  &\abs{\;
    C(T)-\widetilde{C}_{m}(T)
  \;} \\
  &\hspace*{5mm}<
  2^c
  \max
  \left\{
  \abs{\,
    Y(T)-\widetilde{Y}_{m}(T)
  \,},
  \abs{\,
    W(T)-\widetilde{W}_{m}(T)
  \,},
  \abs{\,
    Z(T)-\widetilde{Z}_{m}(T)
  \,}
  \right\}.
\end{split}
\end{equation}
Let $t$ be any computable real number with $T<t<1$.
Then, $W(2,t)=\lim_{m\to\infty}\widetilde{Y}_m(t)$,
where
\begin{equation*}
  \widetilde{Y}_m(t)
  =\sum_{i=1}^{m} \abs{p_i}^2 2^{-\frac{\abs{p_i}}{t}}.
\end{equation*}
The limit value $W(2,t)$ exists since $t<1$.
Let $\beta$ be the infinite binary string such that
$0.\beta$ is the base-two expansion of $W(2,t)\bmod' 1$
with infinitely many ones.

Given $n$ and $\beta_{\lceil Tn/t\rceil}$
(i.e., the first $\lceil Tn/t\rceil$ bits of $\beta$),
one can find $m_0\in\N^+$ such that
\begin{equation*}
  \lceil W(2,t) \rceil-1+0.\beta_{\lceil Tn/t\rceil}
  <\widetilde{Y}_{m_0}(t).
\end{equation*}
This is possible since
$t$ is a computable real number and
$\lceil W(2,t) \rceil-1+0.\beta_{\lceil Tn/t\rceil}<W(2,t)$.
It is then easy to see that
\begin{equation*}
  \sum_{i=m_0+1}^\infty \abs{p_{i}}^2 2^{-\frac{\abs{p_{i}}}{t}} < 2^{-Tn/t}.
\end{equation*}
Raising both sides of this inequality to the power $t/T$ and
using the inequality $a^c+b^c\le (a+b)^c$
for real numbers $a,b>0$ and $c\ge 1$,
\begin{equation*}
  2^{-n}
  >
  \sum_{i=m_0+1}^\infty \abs{p_i}^{2t/T} 2^{-\frac{\abs{p_{i}}}{T}}
  >
  \sum_{i=m_0+1}^\infty \abs{p_{i}}^2 2^{-\frac{\abs{p_{i}}}{T}}
\end{equation*}
and therefore
\begin{equation*}
  2^{-n}
  >
  \sum_{i=m_0+1}^\infty \abs{p_{i}}2^{-\frac{\abs{p_{i}}}{T}}
  \quad\text{and}\quad
  2^{-n}
  >
  \sum_{i=m_0+1}^\infty 2^{-\frac{\abs{p_{i}}}{T}}.
\end{equation*}
It follows that
\begin{equation}\label{eq: ods-dfrC}
\begin{split}
  &\max
  \left\{
  \abs{\,
    W(2,T)-\widetilde{Y}_{m_0}(T)
  \,},
  \abs{\,
    W(1,T)-\widetilde{W}_{m_0}(T)
  \,},
  \abs{\,
    Z(T)-\widetilde{Z}_{m_0}(T)
  \,}
  \right\} \\
  &\hspace*{0mm}<2^{-n}.
\end{split}
\end{equation}
From \eqref{eq: str-dfrC}, \eqref{pfKoatkeC}, \eqref{eq: ods-dfrC}, and
\begin{equation*}
  \abs{\,
    \lfloor C(T) \rfloor + 0.(C(T))_n-C(T)
  \,}
  <2^{-n},
\end{equation*}
it is shown that
\begin{equation*}
  \abs{\;(C(T))_n - g(m_0,n)\;}<2^c+2
\end{equation*}
and therefore
\begin{equation*}
  (C(T))_n
  =g(m_0,n),\,g(m_0,n)\pm 1,\,g(m_0,n)\pm 2,\,\dots,\,g(m_0,n)\pm (2^c+1),
\end{equation*}
where $(C(T))_n$ is regarded as a dyadic integer.
Thus, there are still $2^{c+1}+3$ possibilities of $(C(T))_n$,
so that one needs only $c+2$ bits more in order to determine $(C(T))_n$.

Thus, there exists a partial recursive function
$\Phi\colon \N^+\times\X\times\X\to\X$ such that
\begin{equation*}
  \forall\,n\in\N^+\quad\exists\,s\in\X\quad
  \abs{s}=c+2\;\;\&\;\;
  \Phi(n,\beta_{\lceil Tn/t\rceil},s)=(C(T))_n.
\end{equation*}
It follows from \eqref{eq: fas} that
$H((C(T))_n)
\le |\beta_{\lceil Tn/t\rceil}|+o(n)
\le Tn/t+o(n)$,
which implies that $C(T)$ is $T/t$-compressible.
Since $t$ is any computable real number with $T<t<1$,
it follows that $C(T)$ is $T$-compressible.

By evaluating $C_{n+1}(x)-C_{n}(x)$ for all $x\in(0,1)$
like \eqref{difference-sp},
we can show that
$C_n(x)$ converges uniformly in the wider sense on $(0,1)$ to $C(x)$
as $n\to\infty$.
Hence, we have $C(T)=E'(T)$.

(ii)
It can be shown that
\begin{equation*}
  C_n(1)
  =
  \frac{\ln 2}{2}
  \frac{1}{Z_n(1)^2}
  \sum_{i=1}^n\sum_{j=1}^n
  (\abs{q_i}-\abs{q_j})^2 2^{-\abs{q_{i}}}2^{-\abs{q_{j}}}.
\end{equation*}
By Theorem~\ref{numerator} (i),
$\sum_{j=1}^n(\abs{q_1}-\abs{q_j})^2 2^{-\abs{q_{j}}}$
diverges to $\infty$ as $n\to\infty$.
Therefore
$\sum_{i=1}^n\sum_{j=1}^n
(\abs{q_i}-\abs{q_j})^2 2^{-\abs{q_{i}}}2^{-\abs{q_{j}}}$
also diverges to $\infty$ as $n\to\infty$.
On the other hand,
by Theorem~\ref{crpf} (i),
$Z_n(1)$ converges as $n\to\infty$.
Thus, $C(1)$ diverges to $\infty$,
and the proof is completed.
\end{proof}

Thus,
the theorems in this section show that
the temperature $T$ plays a role as the compression rate
for all the thermodynamic quantities
introduced into algorithmic information theory in this section.
These theorems also show that
the values of the thermodynamic quantities:
partition function, free energy, energy, and statistical mechanical entropy
diverge in the case of $T>1$.
This phenomenon
might
be regarded as
some sort of
phase transition
in statistical mechanics.%
\footnote{
It is still open whether $C(T)$ diverges or not
in the case of $T>1$.
}

\section{Fixed point theorems on compression rate}
\label{fpt}

In this section, we prove the following theorem.

\begin{theorem}[fixed point theorem on compression rate]\label{main}
For every $T\in(0,1)$,
if $Z(T)$ is a computable real number, then the following hold:
\begin{enumerate}
  \item $T$ is right-computable and not left-computable.
  \item $T$ is weakly Chaitin $T$-random and $T$-compressible.
  \item $\lim_{n\to\infty}H(T_n)/n=T$.\qed
\end{enumerate}
\end{theorem}


Theorem~\ref{main} follows immediately from
Theorem~\ref{fpwcTr}, Theorem~\ref{fpTc1}, and Theorem~\ref{fpTc2}
below.
From a purely mathematical point of view,
Theorem~\ref{main} is just a fixed point theorem on compression rate,
where the computability of the value $Z(T)$
gives a sufficient condition for $T$ to be a fixed point
on compression rate.
Note that $Z(T)$ is a monotonically
increasing continuous function on $(0,1)$.
In fact, \cite{T99,T02} showed that
$Z(T)$ is a function of class $C^\infty$ on $(0,1)$.
Thus,
since computable real numbers are dense in $\R$,
we
have the following corollary of Theorem~\ref{main}.

\begin{corollary}
The set $\{T\in(0,1)\mid \lim_{n\to\infty}H(T_n)/n=T\}$
is dense in $[0,1]$.\qed
\end{corollary}

From the point of view of the statistical mechanical interpretation
introduced in the previous section,
Theorem~\ref{main} shows that
the compression rate of temperature is equal to the temperature itself.
Thus,
Theorem~\ref{main} further
confirms
the role of temperature as the compression rate,
which is observed in the previous section.

%
As a first step to prove Theorem~\ref{main},
we prove the following theorem which
gives the weak Chaitin $T$-randomness of $T$
in Theorem~\ref{main}.

\begin{theorem}\label{fpwcTr}
For every $T\in(0,1)$,
if $Z(T)$ is
a right-computable real number,
then $T$ is weakly Chaitin $T$-random.
\end{theorem}

\begin{proof}
Let $p_1,p_2,p_3,\dots$ be a recursive enumeration of the r.e.~set $\Dom U$.
For each $k\in\N^+$,
we define a function $\widetilde{Z}_k\colon(0,1)\to\R$ by
\begin{equation*}
  \widetilde{Z}_k(x)=\sum_{i=1}^{k} 2^{-\frac{\abs{p_i}}{x}}.
\end{equation*}
Then, $\lim_{k\to\infty}\widetilde{Z}_k(x)=Z(x)$ for every $x\in(0,1)$.
On the other hand,
since $Z(T)$ is right-computable,
there exists a total recursive function $g\colon\N^+\to\Q$ such that
$Z(T)\le g(m)$ for all $m\in\N^+$, and
$\lim_{m\to\infty} g(m)=Z(T)$.

We choose any one real number $t$ with $T<t<1$.
For each $i\in\N^+$, using the mean value theorem we see that
\begin{equation*}
  2^{-\frac{\abs{p_i}}{x}}-2^{-\frac{\abs{p_i}}{T}}
  < \frac{\ln 2}{T^2}\abs{p_i}2^{-\frac{\abs{p_i}}{t}}(x-T)
\end{equation*}
for all $x\in(T,t)$.
We choose any one $c\in\N$ with $W(1,t)\ln 2/T^2\le 2^c$.
Here,
the limit value $W(1,t)$ exists by Theorem~\ref{ctr} (i),
since $0<t<1$.
Then, it follows that
\begin{equation}\label{self-contained}
  \widetilde{Z}_k(x)-\widetilde{Z}_k(T)<2^c(x-T)
\end{equation}
for all $k\in\N^+$ and $x\in(T,t)$.

We choose any one $n_0\in\N^+$ such that
$T<0.T_n+2^{-n}<t$ for all $n\ge n_0$.
Such $n_0$ exists since $T<t$ and $\lim_{n\to\infty} 0.T_n+2^{-n}=T$.

Given $T_n$ with $n\ge n_0$,
one can find $k_0,m_0\in\N^+$ such that
\begin{equation*}
  g(m_0)<\widetilde{Z}_{k_0}(0.T_n+2^{-n}).
\end{equation*}
This is possible
from $Z(T)<Z(0.T_n+2^{-n})$,
$\lim_{k\to\infty}\widetilde{Z}_k(0.T_n+2^{-n})=Z(0.T_n+2^{-n})$,
and the properties of $g$.
It follows from $Z(T)\le g(m_0)$
and \eqref{self-contained} that
\begin{equation*}
  \sum_{i=k_0+1}^{\infty} 2^{-\frac{\abs{p_i}}{T}}
  =Z(T)-\widetilde{Z}_{k_0}(T)
  <\widetilde{Z}_{k_0}(0.T_n+2^{-n})-\widetilde{Z}_{k_0}(T)<2^{c-n}.
\end{equation*}
Hence,
for every $i>k_0$,
$2^{-\frac{\abs{p_{i}}}{T}}<2^{c-n}$
and therefore $Tn-Tc<\abs{p_i}$.
Thus,
by calculating the set $\{\>U(p_i)\bigm|i\le k_0\;\}$
and picking any one finite binary string which is not in this set,
one can then obtain an $s\in\X$ such that $Tn-Tc<H(s)$.

Hence, there exists a partial recursive function $\Psi\colon\X\to\X$
such that
$Tn-Tc<H(\Psi(T_n))$
for all $n\ge n_0$.
Using \eqref{eq: k}, there is $c_\Psi\in\N^+$ such that
$H(\Psi(T_n))<H(T_n)+c_\Psi$
for all $n\ge n_0$.
Therefore,
$Tn-Tc-c_\Psi<H(T_n)$
for all $n\ge n_0$.
It follows that $T$ is weakly Chaitin $T$-random.
\end{proof}

\begin{remark}
By elaborating Theorem \ref{crpf} (i),
we can see that
the left-computability of $T$ results in
the weak Chaitin $T$-randomness of $Z(T)$.
On the other hand,
by Theorem \ref{fpwcTr},
the right-computability of $Z(T)$ results in
the weak Chaitin $T$-randomness of $T$.
We can integrate these two extremes into the following form:
For every $T\in(0,1]$,
there exists $c\in\N^+$ such that, for every $n\in\N^+$,
\begin{equation}\label{intermediate1}
  Tn-c\le H(T_n,(Z(T))_n),
\end{equation}
where $H(s,t)$ is defined as $H(<s,t>)$
with any one computable bijection $<s,t>$ from $(s,t)\in \X\times \X$ to $\X$
(see \cite{C75} for the detail of the notion of $H(s,t)$).
In fact,
if $T$ is left-computable,
then we can show that $H((Z(T))_n)=H(T_n,(Z(T))_n)+O(1)$,
and therefore the inequality \eqref{intermediate1} results in
the weak Chaitin $T$-randomness of $Z(T)$.
On the other hand,
if $Z(T)$ is right-computable,
then we can show that $H(T_n)=H(T_n,(Z(T))_n)+O(1)$,
and therefore the inequality \eqref{intermediate1} results in
the weak Chaitin $T$-randomness of $T$.

Note, however, that
the inequality \eqref{intermediate1} is not necessarily tight
except for these two extremes, that is,
the following inequality does not hold:
For every $T\in(0,1]$,
\begin{equation}\label{intermediate2}
  H(T_n,(Z(T))_n)\le Tn+o(n),
\end{equation}
where $o(n)$ depends
on $T$ in addition to $n$.
To see this,
contrarily assume that the inequality \eqref{intermediate2} holds.
Then, by setting $T$ to Chaitin's $\Omega$, we have
$H(\Omega_n)\le H(\Omega_n,(Z(\Omega))_n)+O(1)\le \Omega n+o(n)$.
Since $\Omega<1$,
this contradicts the fact that $\Omega$ is weakly Chaitin random.
Thus, the inequality \eqref{intermediate2} does not hold.
%
\qed
\end{remark}

The following Theorem~\ref{fpTc1} and Theorem~\ref{fpTc2} give
the $T$-compressibility of $T$ in Theorem~\ref{main} together.

\begin{theorem}\label{fpTc1}
For every $T\in(0,1)$,
if $Z(T)$ is
a right-computable real number,
then $T$ is also
a right-computable real number.
\end{theorem}

\begin{proof}
Let $p_1,p_2,p_3,\dots$ be a recursive enumeration of the r.e.~set $\Dom U$.
For each $k\in\N^+$,
we define a function $\widetilde{Z}_k\colon(0,1)\to\R$ by
\begin{equation*}
  \widetilde{Z}_k(x)=\sum_{i=1}^{k} 2^{-\frac{\abs{p_i}}{x}}.
\end{equation*}
Then, $\lim_{k\to\infty}\widetilde{Z}_k(x)=Z(x)$ for every $x\in(0,1)$.
Since $Z(T)$ is right-computable,
there exists a total recursive function $g\colon\N^+\to\Q$ such that
$Z(T)\le g(m)$ for all $m\in\N^+$, and
$\lim_{m\to\infty} g(m)=Z(T)$.
Since $Z(x)$ is an increasing function of $x$,
we see that,
for every $x\in\Q$ with $0<x<1$,
$T<x$ if and only if there are $m,k\in\N^+$ such that
$g(m)<\widetilde{Z}_k(x)$.
Thus, $T$ is right-computable.
This is because
the set
$\{\,(m,n)\,\mid\;m\in\Z\;\&\;n\in\N^+\;\&\;T<m/n\,\}$ is r.e.~if
and only if $T$ is right-computable.
\end{proof}

The converse of Theorem \ref{fpTc1} does not hold.
To see this,
consider an arbitrary computable real number $T\in(0,1)$.
Then,
obviously $T$ is right-computable.
On the other hand,
$Z(T)$ is left-computable and weakly Chaitin $T$-random
by Theorem \ref{crpf} (i).
Thus, $Z(T)$ is not right-computable.

\begin{theorem}\label{fpTc2}
For every $T\in(0,1)$,
if $Z(T)$ is
a left-computable real number
and
$T$ is
a right-computable real number,
then
$T$ is $T$-compressible.
\end{theorem}

\begin{proof}
Let $p_1,p_2,p_3,\dots$ be a recursive enumeration of the r.e.~set $\Dom U$.
For each $k\in\N^+$,
we define a function $\widetilde{Z}_k\colon(0,1)\to\R$ by
\begin{equation*}
  \widetilde{Z}_k(x)=\sum_{i=1}^{k} 2^{-\frac{\abs{p_i}}{x}}.
\end{equation*}
Then, $\lim_{k\to\infty}\widetilde{Z}_k(x)=Z(x)$ for every $x\in(0,1)$.

For each $i\in\N^+$, using the mean value theorem we see that
\begin{equation*}
  2^{-\frac{\abs{p_1}}{t}}-2^{-\frac{\abs{p_1}}{T}}
  > (\ln 2)\abs{p_1}2^{-\frac{\abs{p_1}}{T}}(t-T)
\end{equation*}
for all $t\in(T,1)$.
We choose any one $c\in\N^+$ such that
$(\ln 2)\abs{p_1}2^{-\frac{\abs{p_1}}{T}}\ge 2^{-c}$.
Then, it follows that
\begin{equation}\label{self-contained2}
  \widetilde{Z}_k(t)-\widetilde{Z}_k(T)>2^{-c}(t-T)
\end{equation}
for all $k\in\N^+$ and $t\in(T,1)$.

Since $T$ is a right-computable real number with $T<1$,
there exists a total recursive function $f\colon\N^+\to\Q$ such that
$T<f(l)< 1$ for all $l\in\N^+$, and
$\lim_{l\to\infty} f(l)=T$.
On the other hand,
since $Z(T)$ is left-computable,
there exists a total recursive function $g\colon\N^+\to\Q$ such that
$g(m)\le Z(T)$ for all $m\in\N^+$, and
$\lim_{m\to\infty} g(m)=Z(T)$.
Let $\beta$ be the infinite binary sequence such that $0.\beta$ is
the base-two expansion of $Z(1)$
(i.e.,
Chaitin's
$\Omega$).

Given $n$ and $\beta_{\lceil Tn\rceil}$
(i.e., the first $\lceil Tn\rceil$ bits of $\beta$),
one can find $k_0\in\N^+$ such that
\begin{equation*}
  0.\beta_{\lceil Tn\rceil}<
  \sum_{i=1}^{k_0} 2^{-\abs{p_i}}.
\end{equation*}
It is then easy to see that
\begin{equation*}
  \sum_{i=k_0+1}^{\infty} 2^{-\abs{p_i}}<2^{-Tn}.
\end{equation*}
Using the inequality $a^d+b^d\le (a+b)^d$
for real numbers $a,b>0$ and $d\ge 1$,
it follows that
\begin{equation}\label{tcnew}
  Z(T)-\widetilde{Z}_{k_0}(T)<2^{-n}.
\end{equation}
Note that
$\widetilde{Z}_{k_0}(T)<\widetilde{Z}_{k_0}(f(l))$ for all $l\in\N^+$, and
$\lim_{l\to\infty} \widetilde{Z}_{k_0}(f(l))=\widetilde{Z}_{k_0}(T)$.
Thus, since $\widetilde{Z}_{k_0}(T)<Z(T)$,
one can then find $l_0, m_0\in\N^+$ such that
\begin{equation*}
  \widetilde{Z}_{k_0}(f(l_0))<g(m_0).
\end{equation*}
It follows from \eqref{tcnew} and \eqref{self-contained2}
that
\begin{equation*}
  2^{-n}
  >g(m_0)-\widetilde{Z}_{k_0}(T)
  >\widetilde{Z}_{k_0}(f(l_0))-\widetilde{Z}_{k_0}(T)
  >2^{-c}(f(l_0)-T).
\end{equation*}
Thus, $0<f(l_0)-T<2^{c-n}$.
Let $t_n$ be the first $n$ bits of the base-two expansion of
the rational number $f(l_0)$ with infinitely many zeros.
Then, $\abs{\,f(l_0)-0.t_n\,}<2^{-n}$.
It follows from $\abs{\,T-0.T_n\,}<2^{-n}$ that
$\abs{\,0.T_n-0.t_n\,}<(2^c+2)2^{-n}$.
Hence
\begin{equation*}
  T_n=t_n,\,t_n\pm 1,\,t_n\pm 2,\,\dots,\,t_n\pm (2^c+1),
\end{equation*}
where $T_n$ and $t_n$ are regarded as a dyadic integer.
Thus, there are still $2^{c+1}+3$ possibilities of $T_n$,
so that one needs only $c+2$ bits more in order to determine $T_n$.

Thus, there exists a partial recursive function
$\Phi\colon \N^+\times\X\times\X\to\X$ such that
\begin{equation*}
  \forall\,n\in\N^+\quad\exists\,s\in\X\quad
  \abs{s}=c+2\;\;\&\;\;
  \Phi(n,\beta_{\lceil Tn\rceil},s)=T_n.
\end{equation*}
It follows from \eqref{eq: fas} that
$H(T_n)
\le |\beta_{\lceil Tn\rceil}|+o(n)
\le Tn+o(n)$,
which implies that $T$ is $T$-compressible.
\end{proof}


In a similar manner to the proof of Theorem \ref{main},
we can prove
another version of a fixed point theorem on compression rate as follows.
Here,
the weak Chaitin $T$-randomness is replaced by
the Chaitin $T$-randomness.

\begin{theorem}[fixed point theorem on compression rate II]\label{mainII}
Let $Q$ be a computable real number with $Q>0$.
For every $T\in(0,1)$,
if $W(Q,T)$ is a computable real number, then the following hold:
\begin{enumerate}
  \item $T$ is right-computable and not left-computable.
  \item $T$ is Chaitin $T$-random and $T$-compressible.\qed
\end{enumerate}
\end{theorem}

\begin{remark}
The computability of $Z(T)$ in the premise of Theorem~\ref{main}
can be replaced by the computability of $F(T)$.
On the other hand,
the computability of $W(Q,T)$ in the premise of Theorem~\ref{mainII}
can be replaced by
the computability of
$E(T)$ or $S(T)$.
\qed
\end{remark}

\section{Total statistical mechanical interpretation
of algorithmic information theory:
Physical and informal argument}
\label{smi}

In what follows,
based on a physical argument
we develop a
total
statistical mechanical interpretation
of algorithmic information theory
which
attains
a perfect correspondence to normal statistical mechanics.
In consequence,
we justify the interpretation of $\Omega^D$
as a partition function
and clarify the statistical mechanical meaning of
the thermodynamic quantities introduced
into algorithmic information theory
in Section \ref{tcr}.
In the work \cite{T07},
we developed
a statistical mechanical interpretation of
the noiseless source coding scheme
based on an absolutely optimal instantaneous code
by identifying a microcanonical ensemble in the scheme.
In a similar manner to \cite{T07}
we develop a statistical mechanical interpretation
of algorithmic information theory in what follows.
This can be possible
because
the set $\Dom U$ is prefix-free and therefore
the action of the optimal computer $U$ can be regarded as
an instantaneous code
which is extended over an infinite set.
Note that,
in what follows,
we do not stick to the mathematical strictness of the argument
and
we make an argument
on the same level of mathematical strictness as statistical mechanics
in physics.
We start with some reviews of statistical mechanics.

In statistical mechanics
we consider a quantum system $\mathcal{S}_{\text{total}}$
which consists in a large number of identical quantum subsystems.
Let $N$ be the number of such subsystems.
For example, $N\sim 10^{22}$
for
$1\,\mathrm{cm^3}$
of a gas at room temperature.
We assume here that each quantum subsystem can be distinguishable from others.
Thus, we deal with quantum particles which obey
Maxwell-Boltzmann statistics and not
Bose-Einstein statistics or Fermi-Dirac statistics.
Under this assumption,
we can identify the $i$th quantum subsystem $\mathcal{S}_i$
for each $i=1,\dots,N$.
In quantum mechanics,
any quantum system is described by a quantum state completely.
In statistical mechanics,
among all quantum states,
energy eigenstates are of particular importance.
Any energy eigenstate of each subsystem $\mathcal{S}_i$ can be specified by
a number $n=1,2,3,\dotsc$, called a \textit{quantum number},
where the subsystem in the energy eigenstate specified by $n$ has
the energy $E_n$.
Then, any energy eigenstate of the system $\mathcal{S}_{\text{total}}$
can be specified by an $N$-tuple $(n_1,n_2,\dots,n_N)$ of
quantum numbers.
If the state of the system $\mathcal{S}_{\text{total}}$ is
the energy eigenstate specified by $(n_1,n_2,\dots,n_N)$,
then the state of each subsystem $\mathcal{S}_i$ is
the energy eigenstate specified by $n_i$
and the system $\mathcal{S}_{\text{total}}$
has the energy $E_{n_1}+E_{n_2}+ \dots +E_{n_N}$.
Then,
the fundamental postulate of statistical mechanics,
called \textit{the principle of equal probability},
is stated as follows.
\vspace*{3mm}

\noi
\textbf{The Principle of Equal Probability:}
If the energy of the system $\mathcal{S}_{\text{total}}$ is known to have
a constant value in the range between $E$ and $E+\delta E$,
where $\delta E$ is the indeterminacy
in measurement of the energy of the system $\mathcal{S}_{\text{total}}$,
then the system $\mathcal{S}_{\text{total}}$ is equally likely to be in
any energy eigenstate specified by
$(n_1,n_2,\dots,n_N)$
such that $E\le E_{n_1}+E_{n_2}+ \dots +E_{n_N}\le E+\delta E$.\qed
\vspace*{3mm}

Let $\Theta(E,N)$ be the total number of energy eigenstates
of $\mathcal{S}_{\text{total}}$ specified by $(n_1,n_2,\dots,n_N)$
such that $E\le E_{n_1}+E_{n_2}+ \dots +E_{n_N}\le E+\delta E$.
The above postulate states that
any energy eigenstate of $\mathcal{S}_{\text{total}}$
whose energy lies between $E$ and $E+\delta E$
occurs with the probability $1/\Theta(E,N)$.
This uniform distribution of energy eigenstates
whose energy
lie
between $E$ and $E+\delta E$ is called
a \textit{microcanonical ensemble}.
In statistical mechanics, the \textit{entropy} $S(E,N)$ of
the system $\mathcal{S}_{\text{total}}$ is then defined by
\begin{equation*}
  S(E,N) = k\ln \Theta(E,N),
\end{equation*}
where $k$ is a positive constant, called the \textit{Boltzmann Constant},
and the $\ln$ denotes
the natural logarithm.
The average energy $\varepsilon$ per one subsystem is given by $E/N$.
In a normal case where $\varepsilon$ has a finite value,
the entropy $S(E,N)$ is proportional to $N$.
On the other hand,
the indeterminacy $\delta E$ of the energy
contributes to $S(E,N)$ through the term $k\ln\delta E$,
which can be ignored compared to $N$ unless $\delta E$ is too small.
Thus the magnitude of the indeterminacy $\delta E$ of the energy does
not matter to the value of the entropy $S(E,N)$ unless it is too small.
The \textit{temperature} $T(E,N)$ of the system $\mathcal{S}_{\text{total}}$
is defined by
\begin{equation*}
  \frac{1}{T(E,N)} = \frac{\partial S}{\partial E}(E,N).
\end{equation*}
Thus the temperature is a function of $E$ and $N$.

Now we give a statistical mechanical interpretation
to algorithmic information theory.
As considered in \cite{C75},
think
of the optimal computer $U$ as decoding equipment at the receiving end of
a noiseless binary communication channel.
Regard
its programs (i.e., finite binary strings in $\Dom U$)
as codewords and regard the result
of the computation by $U$,
which is a finite binary string,
as a decoded ``symbol.''
Since $\Dom U$ is a prefix-free set,
such codewords form what is called an ``instantaneous code,''
so that successive symbols sent through the channel
in the form of concatenation of codewords can be separated.

For establishing the statistical mechanical interpretation
of algorithmic information theory,
we assume that
the infinite binary string sent through the channel
is generated by infinitely repeated tosses of a fair coin.
Under this assumption,
the success probability of decoding one symbol
is equal to Chaitin's halting probability $\Omega$,
and the probability of getting a finite binary string $s$
as the first decoded symbol is equal to $P(s)$.
Hereafter
the infinite binary string sent through the channel
is referred to as
\textit{the channel infinite string}.
For each $r\in\X$,
let $Q(r)$ be the probability that
the channel infinite string has the prefix $r$.
It follows that $Q(r)=2^{-\abs{r}}$.
Thus, the channel infinite string is the random variable
drawn according to Lebesgue measure on $\XI$.

Let $N$ be a large number, say $N\sim 10^{22}$.
We relate algorithmic information theory to
the statistical mechanics reviewed above
in the following manner.
Among all infinite binary strings,
consider infinite binary strings of the form $p_1 p_2 \dotsm p_N \alpha$
with $p_1,p_2,\dots,p_N\in\Dom U$ and $\alpha\in\XI$.
For each $i$,
the $i$th slot fed by $p_i$ corresponds to
the $i$th quantum subsystem $\mathcal{S}_i$.
On the other hand,
the ordered sequence of the $1$st slot, the $2$nd slot, $\dotsc$,
and the $N$th slot
corresponds to the quantum system $\mathcal{S}_{\text{total}}$.
We relate a codeword $p\in\Dom U$ to an energy eigenstate of a subsystem,
and
relate a codeword length $\abs{p}$ to an energy $E_n$ of
the energy eigenstate of the subsystem.
Then, a finite binary string $p_1 \dotsm p_N$ corresponds to
an energy eigenstate of $\mathcal{S}_{\text{total}}$
specified by $(n_1,\dots,n_N)$.
Thus, $\abs{p_1}+\dots+\abs{p_N}=\abs{p_1 \dotsm p_N}$ corresponds to
the energy $E_{n_1}+ \dots +E_{n_N}$ of
the energy eigenstate of $\mathcal{S}_{\text{total}}$.

We define a subset $C(L,N)$ of $\X$ as
the set of all finite binary strings of the form
$p_1 \dotsm p_N$ with $p_i\in\Dom U$
whose total length
$\abs{p_1 \dotsm p_N}$
lie between $L$ and $L+\delta L$.
Then, $\Theta(L,N)$ is defined as
the cardinality of $C(L,N)$.
Therefore, $\Theta(L,N)$ is the total number of
all concatenations of $N$ codewords
whose total length lie between $L$ and $L+\delta L$.
We can see that
if $p_1 \dotsm p_N\in C(L,N)$,
then $2^{-(L+\delta L)}\le Q(p_1 \dotsm p_N) \le 2^{-L}$.
Thus,
all concatenations $p_1 \dotsm p_N\in C(L,N)$ of $N$ codewords occur
in a prefix of the channel infinite string
with the same probability $2^{-L}$.
Note here that we care nothing about the magnitude of $\delta L$,
as in the case of statistical mechanics.
Thus, the following principle,
called \textit{the principle of equal conditional probability},
holds.
\vspace*{3mm}

\noi
\textbf{The Principle of Equal Conditional Probability:}
Given that a concatenation of $N$ codewords of total length $L$
occurs in a prefix of the channel infinite string,
all such concatenations occur
with the same probability $1/\Theta(L,N)$.\qed
\vspace*{3mm}

We introduce a microcanonical ensemble
into algorithmic information theory in this manner.
Thus, we can develop a certain sort of statistical mechanics
on algorithmic information theory.
Note that,
in statistical mechanics,
the principle of equal probability
is just a conjecture which is not yet proved completely
in a realistic physical system.
On the other hand,
in our statistical mechanical interpretation
of algorithmic information theory,
the principle of equal conditional probability
is automatically satisfied.

The \textit{statistical mechanical entropy} $S(L,N)$ is defined by
\begin{equation}\label{def-smentropy}
  S(L,N)=\log_2 \Theta(L,N).
\end{equation}
The \textit{temperature} $T(L,N)$ is then defined by
\begin{equation}\label{def-temperature}
  \frac{1}{T(L,N)} = \frac{\partial S}{\partial L}(L,N).
\end{equation}
Thus, the temperature is a function of $L$ and $N$.

According to the theoretical development of
equilibrium statistical mechanics,%
\footnote{
We follow the argument of Section 16-1 of Callen \cite{C85}
in particular.
}
we can introduce a canonical ensemble
into algorithmic information theory in the following manner.
We investigate the probability distribution of
the left-most codeword $p_1$ of the channel infinite string,
given that a concatenation of $N$ codewords of total length $L$
occurs in a prefix of the channel infinite string.
For each $p\in\X$,
let $R(p)$ be the probability that
the left-most codeword of the channel infinite string is $p$,
given that a concatenation of $N$ codewords of total length $L$
occurs in a prefix of the channel infinite string.
Based on the principle of equal conditional probability,
it can be shown that
\begin{equation*}
  R(p)=\frac{\Theta(L-\abs{p},N-1)}{\Theta(L,N)}.
\end{equation*}
From the general definition \eqref{def-smentropy} of
statistical mechanical entropy,
we have
\begin{equation}\label{r2ss}
  R(p)=2^{S(L-\abs{p},N-1)-S(L,N)}.
\end{equation}

Let $E(L,N)$ be the expected length of the left-most codeword of
the channel infinite string,
given that a concatenation of $N$ codewords of total length $L$
occurs in a prefix of the channel infinite string.
Then, the following equality is expected to hold:
\begin{equation}\label{additivity}
  S(L,N)=S(E(L,N),1)+S(L-E(L,N),N-1).
\end{equation}
Here, the term $S(L,N)$ in the left-hand side
denotes the statistical mechanical entropy of
the whole concatenation
of $N$ codewords of total length $L$.
On the other hand,
the first term $S(E(L,N),1)$ in the right-hand side denotes
the statistical mechanical entropy of
the left-most codeword of
the concatenation
of $N$ codewords of total length $L$
while the second term $S(L-E(L,N),N-1)$ in the right-hand side denotes
the statistical mechanical entropy of
the remaining $N-1$ codewords of
the concatenation
of $N$ codewords of total length $L$.
Thus, the equality \eqref{additivity} represents
the additivity of the statistical mechanical entropy.
We assume here that the equality \eqref{additivity} holds.

By expanding $S(L-\abs{p},N-1)$ around the equilibrium point
$L-E(L,N)$,
we have
\begin{align}
  &S(L-\abs{p},N-1) \nonumber \\
  =&S(L-E(L,N)+E(L,N)-\abs{p},N-1) \nonumber \\
  =&S(L-E(L,N),N-1)
   +\frac{\partial S}{\partial L}(L-E(L,N),N-1)(E(L,N)-\abs{p}).
   \label{taylor}
\end{align}
Here, we ignore the higher order terms than the first order.
Since $N\gg 1$ and $L\gg E(L,N)$,
using the definition \eqref{def-temperature} of temperature
we have
\begin{equation}\label{approximation}
  \frac{\partial S}{\partial L}(L-E(L,N),N-1)
  =\frac{\partial S}{\partial L}(L,N)
  =\frac{1}{T(L,N)}.
\end{equation}
Hence, by \eqref{taylor} and \eqref{approximation},
we have
\begin{equation}\label{step}
  S(L-\abs{p},N-1)
  =S(L-E(L,N),N-1)+\frac{1}{T(L,N)}(E(L,N)-\abs{p}).
\end{equation}

Thus, using \eqref{r2ss}, \eqref{additivity},
and \eqref{step},
we obtain
\begin{equation*}
  R(p)
  =2^{\frac{E(L,N)-T(L,N)S(E(L,N),1)}{T(L,N)}}
   2^{-\frac{\abs{p}}{T(L,N)}}.
\end{equation*}
Then,
according to statistical mechanics
we define the \textit{free energy} $F(L,N)$
of the left-most codeword of
the concatenation
of $N$ codewords of total length $L$
by
\begin{equation}\label{def-free-energy-nm}
  F(L,N)=E(L,N)-T(L,N)S(E(L,N),1).
\end{equation}
It follows that
\begin{equation}\label{RFTpT}
  R(p)
  =2^{\frac{F(L,N)}{T(L,N)}}
   2^{-\frac{\abs{p}}{T(L,N)}}.
\end{equation}
Using $\sum_{p\in\Dom U}R(p)=1$,
we can show that, for any $p\in\Dom U$,
\begin{equation}\label{RZpT}
  R(p)=\frac{1}{Z(T(L,N))}2^{-\frac{\abs{p}}{T(L,N)}},
\end{equation}
where $Z(T)$ is defined by
\begin{equation}\label{partition_function_pa}
  Z(T)=
  \sum_{p\in\Dom U}2^{-\frac{\abs{p}}{T}}
  \qquad ( T > 0 ).
\end{equation}
$Z(T)$ is called the \textit{partition function}
(of the left-most codeword of the channel infinite string).
Thus, in our statistical mechanical interpretation
of algorithmic information theory,
the partition function $Z(T)$ has exactly the same form as $\Omega^D$.
The distribution in the form of $R(p)$ is called
a \textit{canonical ensemble} in statistical mechanics.

Then, using \eqref{RFTpT} and \eqref{RZpT},
$F(L,N)$ is calculated as
\begin{equation}\label{fft}
  F(L,N)=F(T(L,N)),
\end{equation}
where $F(T)$ is defined by
\begin{equation}\label{free_energy_pa}
  F(T)=-T\log_2 Z(T)
  \qquad ( T > 0 ).
\end{equation}

On the other hand,
from the definition of $R(p)$,
$E(N,L)$ is calculated
as
\begin{equation*}
  E(N,L)=\sum_{p\in\Dom U} \abs{p}R(p).
\end{equation*}
Thus, we have
\begin{equation}\label{EET}
  E(L,N)=E(T(L,N)),
\end{equation}
where $E(T)$ is defined by
\begin{equation}\label{energy_pa}
  E(T)=\frac{1}{Z(T)}
  \sum_{p\in\Dom U} \abs{p}2^{-\frac{\abs{p}}{T}}
  \qquad ( T > 0 ).
\end{equation}

Then,
using \eqref{def-free-energy-nm}, \eqref{fft},
\eqref{free_energy_pa}, and \eqref{EET},
the statistical mechanical entropy $S(E(L,N),1)$
of the left-most codeword of
the concatenation
of $N$ codewords of total length $L$
is calculated as
\begin{equation*}
  S(E(L,N),1)=S(T(L,N)),
\end{equation*}
where $S(T)$ is defined by
\begin{equation}\label{entropy_pa}
  S(T)=\frac{1}{T}E(T)+\log_2 Z(T)
  \qquad ( T > 0 ).
\end{equation}
Note that
the statistical mechanical entropy $S(E(L,N),1)$ coincides with
the \textit{Shannon entropy}
\begin{equation*}
  -\sum_{p\in\Dom U} R(p)\log_2 R(p)
\end{equation*}
of the distribution $R(p)$.

Finally,
the \textit{specific heat} $C(T)$
of the left-most codeword of
the channel infinite string
is
defined by
\begin{equation}\label{specific_heat_pa}
  C(T)=E'(T)
  \qquad ( T > 0 ),
\end{equation}
where $E'(T)$ is the derived function of $E(T)$.

Thus,
a statistical mechanical interpretation
of algorithmic information theory
can be established,
based on a physical
argument.
We can check that
the formulas in this argument:
the partition function \eqref{partition_function_pa},
the free energy \eqref{free_energy_pa},
the expected length of the left-most codeword
\eqref{energy_pa},
the statistical mechanical entropy \eqref{entropy_pa}, and
the specific heat \eqref{specific_heat_pa}
correspond to
the definitions in Section \ref{tcr}:
Definition \ref{def-partition-function},
Definition \ref{def-free-energy},
Definition \ref{def-energy},
Definition \ref{def-entropy}, and
Definition \ref{def-specific-heat},
respectively.
Thus,
the statistical mechanical meaning of
the notion of thermodynamic quantities
introduced in Section \ref{tcr}
into algorithmic information theory
is clarified
by this argument.


\section{Concluding remarks}
\label{conclusion}

In this paper,
we have developed a statistical mechanical interpretation of
algorithmic information theory
by introducing
the notion of thermodynamic quantities
into algorithmic information theory
and investigating their properties
from the point of view of algorithmic randomness.
As a result,
we have discovered that,
in the interpretation,
the temperature plays a role as
the compression rate of
all
these thermodynamic quantities,
which include the temperature itself.
Thus,
in particular,
we have obtained
fixed point theorems on compression rate,
which reflect this self-referential nature of
the compression rate of the temperature.
In the last part of this paper,
we have also developed
a total statistical mechanical interpretation of
algorithmic information theory,
which realizes a perfect correspondence to normal statistical mechanics
and motives the above introduction of the thermodynamic quantities
into algorithmic information theory.
However,
the argument used in the total statistical mechanical interpretation
is on the same level of mathematical strictness as statistical mechanics.
Thus, we try to make the argument
a mathematically rigorous form in a future study.
This effort might stimulate a further unexpected development
of the research of this line.

\section*{Acknowledgments}

The author is grateful to Prof.~Shigeo Tsujii for the financial supports.


\end{document}